\documentclass[11pt,a4paper]{article}

\usepackage{graphicx,eurosym}
\usepackage{float}
\usepackage{afterpage}
\usepackage{epstopdf}
\usepackage{epsfig,cite}
\usepackage{amssymb}
\usepackage{amsmath}
\usepackage{dsfont}
\usepackage{multirow}
\usepackage{url}
\usepackage[colorlinks=false]{hyperref}
\usepackage[ascii]{inputenc}
\usepackage[T1]{fontenc}
\usepackage{amsmath}
\usepackage{amssymb,amsfonts,textcomp}
\usepackage{color}
\usepackage{url}
\usepackage{hyperref}
\usepackage{array}
\usepackage{supertabular}
\usepackage{hhline}
\usepackage{hyperref}
\hypersetup{colorlinks=true, linkcolor=blue, citecolor=blue, filecolor=blue, urlcolor=blue}
\usepackage{graphicx}
\newcommand\textsubscript[1]{\ensuremath{{}_{\text{#1}}}}
\newcommand\textstylefootnotereference[1]{\textsuperscript{#1}}
\makeatletter
\newcommand\arraybslash{\let\\\@arraycr}
\makeatother
\makeatletter
\newcommand\ps@Standard{
  \renewcommand\@oddhead{}
  \renewcommand\@evenhead{\@oddhead}
  \renewcommand\@oddfoot{}
  \renewcommand\@evenfoot{}
  \renewcommand\thepage{\arabic{page}}
}
\makeatother
\newcommand{\be}{\begin{equation}}
\newcommand{\ee}{\end{equation}}
\newcommand{\bea}{\begin{eqnarray}}
\newcommand{\eea}{\end{eqnarray}}
\newcommand{\bi}{\begin{itemize}}
\newcommand{\ei}{\end{itemize}}
\newcommand{\ben}{\begin{enumerate}}
\newcommand{\een}{\end{enumerate}}

\def\frac#1#2{{{#1}\over {#2}}}
\def\gsim{\mathrel{\rlap{\lower4pt\hbox{\hskip1pt$\sim$}}
    \raise1pt\hbox{$>$}}}         
\def\lsim{\mathrel{\rlap{\lower4pt\hbox{\hskip1pt$\sim$}}
    \raise1pt\hbox{$<$}}}         

\newcommand{\draft}[1]{}

\def\beq{\begin{equation}}  
\def\eeq{\end{equation}}  

\def \n0{N_j^{(0)}}

\def\lapprox{\lower .7ex\hbox{$\;\stackrel{\textstyle <}{\sim}\;$}}
\def\gapprox{\lower .7ex\hbox{$\;\stackrel{\textstyle >}{\sim}\;$}}

\begin{document}
\begin{flushright}
TIF-UNIMI-2016-01\\
\end{flushright}

\vspace{0.4cm}

\begin{center} 
  {\bf{\Large Forecasting the Socio-Economic  Impact of the Large
      Hadron Collider: a Cost-Benefit Analysis to 2025 and Beyond}}
\vspace{.7cm}

Massimo Florio$^1$, Stefano Forte$^2$, 
 and Emanuela Sirtori$^3$ 

\vspace{.3cm}
{\it~$^1$ Dipartimento di Economia, Management e Metodi Quantitativi,\\
Universit\`a di Milano , via Conservatorio 7, I-20122 Milano, Italy\\
~$^2$ TIF Lab, Dipartimento di Fisica, Universit\`a di Milano and\\
INFN, Sezione di Milano, Via Celoria 16, I-20133 Milano, Italy\\
~$^3$ CSIL, Centre for Industrial Studies\\
Corso Monforte 15,  I-20122 Milano, Italy}
\end{center}   

{\centering
\textbf{Abstract}
\par}

In this paper we develop a cost-benefit analysis of a major research
infrastructure, the Large Hadron Collider (LHC), the highest-energy
accelerator in the world, currently operating at CERN. We show that the
evaluation of benefits can be made quantitative by estimating their
welfare effects on different types of agents. Four classes of direct
benefits are identified, according to the main social groups involved:
(a) scientists; (b) students and young researchers; (c) firms in the
procurement chain and other organizations; (d) the general public,
including onsite and website visitors and other media users. These
benefits are respectively related to the knowledge output of
scientists; human capital formation; technological spillovers; and
direct cultural effects for the general public. Welfare effects for
taxpayers can also be estimated by the contingent valuation of the
willingness to pay for a pure public good for which there is no
specific direct use (i.e., as non-use value). Using a Monte Carlo
approach, we estimate the conditional probability distribution of costs
and benefits for the LHC from 1993 until its planned decommissioning in
2025, assuming a range of values for some critical stochastic
variables. We conservatively estimate that there is around a 90\%
probability that benefits exceed costs, with an expected net present
value of about 2.9 billion euro, not considering the unpredictable
applications of scientific discovery.
\clearpage

\section{Introduction}

Cost-benefit analysis (CBA) is widely used by governments and economists
to evaluate the socio-economic impact of investment projects; it
requires the forecasting of inputs, outputs, and their marginal social
values (MSVs) in order to determine the expected net present value
($NPV$) of a project. CBA theory is reviewed for example by Dr\`eze and
Stern 1987, 1990, Johansson 1991, Boardman et al. 2006, Florio 2014,
and Johansson and Kristr\"om 2015. In this framework, a project is
desirable if its social benefits exceed costs over time. This approach
is well developed for conventional infrastructure and is supported for
example by the World Bank, the European Commission, the European
Investment Bank, the OECD, and other national and international
institutions (Baum and Tolbert, 1985 and World Bank 2010; European
Commission 2014, European Investment Bank 2013, and OECD 2015; for the
WHO, see Hutton and Rehfuess, 2006). 

Until now, the application of CBA to research infrastructure (RI) has
been hindered, however, by claims that the unpredictability of future
economic benefits of science creates a difficulty for any quantitative
forecasts. For example OECD 2014 (p.12), in a recent study of the
social impact of CERN, states that a qualitative approach is preferred
because of possible criticism of quantitative methods. In a survey of
past experience, Martin and Tang (2007, p.15)---while noting
substantial advances in empirical analysis of the different channels
through which research expenditures spill over to society---conclude
that it is impossible to compare the different channels of propagation
of the social benefits of science, or to provide {\textquotedblleft}a
quantitative answer to the question of how the overall level of
benefits from basic research compares with the level of public
investment in such research.{\textquotedblright} They suggest that
quantitative forecasts would lead to underestimation of the benefits,
and cite Feller et al. 2002, who report that according to survey data,
{\textquotedblleft}firms investing in university research do not
attempt to make any cost-benefit analysis of this investment on the
grounds that it would be too complex and costly.{\textquotedblright} 

\ \ We acknowledge that CBA of research infrastructure is complex and
that there is a risk of underestimation of benefits. Nevertheless,
given the importance and the increasing cost of science, the potential
advantages for decision-makers of exploring new ways to measure and
compare social benefits and costs of large-scale research
infrastructure cannot be exaggerated. 

\ \ What follows is an application of the CBA framework developed by
Florio and Sirtori (2015), and Florio et al (2016) and should be seen
as a way to explore its feasibility in practice. There are two
important caveats. First, we are not claiming that decisions on funding
scientific projects should be based exclusively on their measurable
socio-economic impact, as there clearly are several other
considerations at stake (the scientific case itself, strategic and
ethical issues, etc.). Second, our approach is conservative, because it
deliberately leaves out several qualitative evaluation issues. In
particular, a novelty of our approach is to make a sharp distinction
between what is measurable and what is not measurable and to focus
exclusively on the former. We shall show that even leaving aside what
cannot be predicted in quantitative terms, including the long-term
effect of a discovery, a proper CBA model can still be applied to
large-scale research infrastructure with interesting empirical
findings.

\ \ The Large Hadron Collider (LHC), our case study, is the biggest
experimental machine in the world (CERN 2009). This, arguably, is a
stringent test of the practical applicability of the Florio and Sirtori
(2015) methodology, because of the very large scale of the project, its
long time horizon, its peculiar international management, and finally
because the LHC{\textquoteright}s physics is basic science, at present
without any predictable economic application. 

\ \ The structure of the paper is the following: in the next section we
briefly present the object of our analysis, the LHC, and why it poses a
challenge for CBA; in section 3 we introduce our CBA model; section 4
briefly describes data sources and methods; section 5 is about
estimation of costs; section 6 deals with the direct value of
publications to scientists; section 7 presents the social benefits of
technological externalities; section 8 considers the human capital
effects of the LHC; section 9 offers a forecast of the cultural
effects; section 10 enlarges the scope of the analysis to non-use
benefits; and section 11 concludes.

\section{The Large Hadron Collider}

The LHC is currently the largest particle accelerator in the world. A
particle accelerator is a device in which particles (protons and atomic
nuclei, in the case of the LHC) are accelerated and made to collide
with a target or with each other, with the goal of studying the
structure of matter. Particles are accelerated by subjecting them to
electric fields and are collimated into focused beams by magnetic
fields. Particle beams travel in a pipe in which a vacuum has been
established and are brought to collide in experimental areas in which
the debris from the collisions is accurately measured by devices called
detectors, which allow for an accurate reconstruction of what has
happened during the collision. 

The main goal of the LHC is to study the precise nature of the forces
that govern fundamental interactions at the shortest distances that are
currently accessible, which requires the colliding particles to hit
each other at the highest possible energy. 

In operation since 2009, a first goal was reached with the discovery in
2012 of the {\textquotedblleft}Higgs boson,{\textquotedblright} at the
time the only major missing piece of information in the existing theory
of fundamental interactions. Current research involves both
investigating the properties of the newly discovered Higgs boson and
searches for deviations from the current theory, which is believed to
be incomplete, and is foreseen to continue for at least about another
decade.

The LHC was built by the European Organization for Nuclear Research
(CERN). Construction work lasted from 1993 to 2008. The LHC is the
largest element of a chain of machines that accelerate particles to
increasingly higher energies---the CERN accelerator complex. The
accelerator complex is developed, maintained, and operated by CERN.
This facility is exploited by the experimental Collaborations that
perform experiments in the areas where collisions occur. Each
experiment is based on a detector, designed, built, and operated by a
Collaboration that involves both the participation of CERN and of
scientists from a number of institutions (universities and research
labs) from several countries. Four main experiments exploit LHC
collisions; the two largest ones both involve several thousand
scientists from several hundred institutions in almost fifty countries.
The corresponding detectors are roughly the size of a ten-story
building. When observing particle collisions, the four experiments
produce about 1 GB of data per second, which are either analyzed inside
by LHC Collaborations or sent to a number of other computer centers
around the world, connected through the worldwide LHC computer grid.

This context is particularly challenging for cost-benefit analysis for
several reasons. First, this is a very large infrastructure by all
measures: number of people involved, physical size, cost. Also, it has
an especially complicated structure due to the intricate interplay of
accelerator and detectors in the experimental Collaborations between
the host laboratory (CERN) and its participating institutions, with the
large number of countries and different kinds of organizations involved
(universities, research labs, national academies). This poses difficult
cost apportionment and aggregation issues when attempting to estimate
costs and benefits.

Second, the life-span (both past and future) of the facility is quite
long: this requires both retrospective evaluation and appraisal
techniques, since capital costs for the LHC were incurred starting from
1993 and the generation of both operating costs and benefits are
expected to continue for some years in the future.

Third, because the LHC is an infrastructure for fundamental research,
the evaluation of its benefits cannot be based on an estimate of the
applications of its discoveries.

In view of all this, we will argue that the application of a CBA model
to the LHC is a form of validation of the model itself, in that the
successful application of the model in this context guarantees that the
model will be able to handle more conventional or simpler situations,
such as infrastructure of a more applied nature, of a smaller scale,
and with a simpler legal and organizational structure. 

\section{The model}

In general, an investment project passes a CBA test if $NPV>0$. If  ${B}_{{t}_{i}}$\textsubscript{ }and 
${C}_{{t}_{i}}$\textsubscript{ }are respectively benefits and costs
incurred at various times  ${t}_{i}$,
\begin{equation}\label{eq:npvdef}
 NPV= \sum_i \frac{B_{t_i}-C_{t_i}}{(1+r)^{t_i}},
\end{equation}
with $r$ the social discount rate, needed to convert a future
value at $t$ in terms of a reference level at $t=0$. We
do not explicitly include an expectation operator in this notation, but
all the variables should be considered as stochastic and are taken here
at their mean values, given their probability distribution functions.
In turn, $B$ and $C$ include $i=1,2,\dots,I$ input and output flows,
each occurring at time $ t=0,1,2,\dots,T$ and valued by shadow prices reflecting
their MSVs (Dr\`eze and Stern 1987, Florio 2014). 

In order to address the evaluation problem quantitatively, we build on
the model developed by Florio and Sirtori (2015), and Florio et al
(2016) to which the reader can refer for details of the approach,
including a review of previous related literature. Borrowing some ideas
from environmental CBA (Johansson 1995, Johansson and Kristr\"om 2015,
Pearce et al. 2006, Atkinson and Mourato 2008), Florio and Sirtori
(2015) break down the $NPV$ of an RI ( $NPV_{RI}$)
into two parts: net use-benefits, i.e., net benefits to those who
{\textquotedblleft}use{\textquotedblright}{\textquoteright} in
different ways the services delivered by the LHC (
$NPV_u$); and the present non-use value of the LHC, i.e.,
its value for people who currently do not use its services, but who
derive utility by just knowing that new science is created (
${B}_{n}$), such that:

\begin{equation}\label{eq:npv}
NPV_{RI}=NPV_u+B_n=\left(PV_{B_u}-PV_{C_u}\right)+\left(QOV_0-EXV_0\right).
\end{equation}

The first term on the r.h.s.,  $NPV_u$, is the time
discounted sum of (negative) capital and operating costs ($PV_{C_u}$), and the
economic value of benefits ($PV_{B_u}$), in turn determined by asking who the direct
beneficiaries of the RI are. It is an intertemporal value, i.e., it has
the structure of Eq.~(\ref{eq:npv}). The  $B_n$ term captures two types of
non-use values related to future discoveries: their quasi-option value
( $QOV_0$) (Arrow and Fisher 1974), which is related to
any future, but unpredictable economic benefit of new knowledge; and an
existence value (Johansson and Kristr\"om, 2015 p.25), which is related
to pure new knowledge per se ( $EXV_0$).  $B_n$ is an
instantaneous value, i.e., it refers to time $t = 0$.

In order to determine $NPV_u$, here, as in Florio and
Sirtori (2015), we ask first who the direct beneficiaries of an RI are
and thus identify four classes of benefits, related to social groups:
(a) scientists; (b) students and post-docs; (c) firms in the supply
chain of the LHC and other organizations; (d) general public exposed to
LHC outreach activities. Starting from (a), the ability to publish new
research findings is the core benefit to scientists, both project
insiders and outsiders ($SC$); (b) benefits for students and
post-docs in terms of future salary and job opportunities arise from
human capital formation, because of the skills gained and the
reputational effects of their training experience at the RI
($HC$); technological externalities (c) are benefits to firms
both in the supply chain of the project procurement and to external
firms involved in technology transfer and also to other organizations
and businesses that save costs because of spillovers from the research
infrastructure activities ($T$); (d) cultural effects are
enjoyed by outreach beneficiaries, including those visiting the
facilities and related exhibitions elsewhere, those who access websites
and social media, and those who enjoy the general media exposure of LHC
activities and discoveries ($CU$). Costs are determined as the
sum of the economic value of capital ($K$), labor cost of
scientists ($LS$) and other staff ($LO$), and operating
costs ($O$)\footnote{ In principle, negative externalities and
other non-market related effects should also be considered. In the case
of the LHC, we assume that these are either negligible (there is no
pollution arising from the infrastructure, as it is mostly located
around one hundred meters underground, and is carefully inspected for
radioprotection) or unpredictable to date (external impact of major
accidents or decommissioning costs, as the latter would depend on
technical decisions possibly taken beyond 2040).}. 

Of the two components of the non-use value  ${B}_{n}$, the quasi-option
value $QOV_0$ is very uncertain. In
principle it would include serendipity effects or any other long-term
impacts that cannot be predicted now in terms of probabilities (Knight
1964). The standard definition of QOV in earlier literature is related
to irreversibility (the fact that certain projects definitively change
a site or some stock of resources), uncertainty of demand for
alternative projects (for which probabilities can, however, be
guessed), and the value of delaying a decision to acquire additional
information. In fact, Johansson (1995) and Pearce et al. (2006) suggest
that QOV should not be included among the benefits, but considered
separately as an information issue. Moreover, the LHC is already
running and our CBA is not fully ex-ante, hence there is no scope now
to evaluate the option to delay the start-up or other technological
options (discussed by Schopper 2009). While in certain domains and for
certain research projects it might be possible in principle to compute
a QOV, this does not seem appropriate for the LHC. Nobody can say
ex-ante what is the use-benefit for society of (possibly) discovering
supersymmetric particles or the possible direct uses of the knowledge
that the Higgs boson exists. Hence the social cost of delaying such
discovery is fully unknown; also unknown is the direct benefit of
having generated such knowledge before it would have been possible
otherwise. 

We thus take $QOV_0$ as not measurable
for the LHC; we just assume that it is non-negative and we set it to
zero. This is the main conservative assumption of our method of
computing the $NPV$. We suspect that our assumption implies an
underestimation of the social benefits, but it also has the advantage
of removing an immeasurable object from the analysis, an issue that
would otherwise be a source of purely speculative guesses. 

 However, the existence value $EXV_0$ is
measurable in principle. It is the social benefit of knowledge per se,
without any direct use. This is a pure public good, not conceptually
different from other Samuelsonian (non rival, non excludable) global
public goods, such as the integral conservation of natural habitats, of
biodiversity, or of cultural heritage, considered separately from any
direct economic exploitation of the protected goods. In environmental
CBA, the existence value is the benefit of preserving something known
to exist (European Commission 2014, Pearce et al. 2006); in the Florio
and Sirtori (2015) framework, it is the benefit of knowing that
something exists. 

The standard welfare economics theory for a pure public good is that it
is socially optimal to provide such a good when the sum of the
willingness to pay (WTP) by taxpayers is equal to the social cost of
provision (Myles 1995, Johansson and Krist\"om 2015). Thus, EXV can be
proxied by an empirical estimation of WTP of knowledge per se by the
general public. We cannot exclude the possibility that in eliciting the
WTP there may be a mixture of EXV and perceived QOV, for which however
the information is not available to the respondents (see Catalano et
al. 2016). 

In sum, our social accounting\footnote{ For further details on this
model, see Florio and Sirtori (2015).} is
\begin{equation}\label{eq:npvres}
{  NPV}= \sum_i \frac{\left({  SC}_{t_i}+{  TE}_{t_i}+{  HC}_{t_i}+{ 
   CU}_{t_i}\right)-\left({  K}_{t_i}+{  LS}_{t_i}+{ 
    LO}_{t_i}+{  O}_{t_i}\right)}{(1+r)^{t_i}}+{  EXV}_0.
\end{equation}
Each variable in Eq.~(\ref{eq:npvres}) is split into several contributions determined
by other variables (e.g., scientists{\textquoteright} salaries on the
cost side, or additional profits of RI suppliers on the benefit side,
etc.), and it is treated as stochastic, as is further explained in the
next section.

\section{Data and methods}

The empirical analysis supporting the evaluation of the socio-economic
impact of the LHC is supported by several sources of data, which are
reported in detail below in the presentation of each cost or benefit
item. The main categories are: (a) accounting data and expert analysis
of capital and operating expenditures, including in-kind contributions;
(b) scientometric data to estimate trajectories of publications and
their impact in a specific domain; (c) firms{\textquoteright} survey
data on technological spillovers expressed in terms of increased sales
and cost savings, or increased profits; expert analysis of the
technological content of procurement; company accounting data for
industries involved in procurement; and expert analysis of the cost
savings or other quantifiable effects of open source software or other
technological spillovers; (d) survey data and other statistical
evidence of the expected or ex-post effects on salaries of former
students and early career scientists; (e) statistics about on-site
visitors, web access, use of social media, exposure to traditional
media, and data on travel costs, opportunity costs of time, and other
information related to cultural effects; (f) contingent valuation data
through survey of samples of potential taxpayers about their WTP for
potential discoveries related to a specific project. 

Financial costs (interest rates arising from borrowing, taxes, and other
cash transfers) have not been included, as they are monetary
transactions that do not create value within the society at the
aggregate level. These are not welfare effects, as stated by CBA
guidelines adopted by national and international organizations. They
would be part of a financial analysis, which is not our objective. We
have also excluded the opportunity cost of public funds, as this would
be related to a marginal effect of distortionary taxation, which for
international grants (the way the LHC is funded) is usually not
considered. Moreover, there has been no special grant to CERN by the
Member States for the purpose of building and operating the LHC, which
has been funded by the regular CERN budget, loans, and---for the
detectors--- by a very large number of contributions, including
in-kind, by CERN member and non-member states (on both issues, see
European Commission 2014).

The data collection required in-depth interviews of more than 1500
people, including PhD candidates and former LHC students,
non-LHC-related students in five European universities, experts at CERN
and elsewhere, company managers and {\textquotedblleft}head
hunters{\textquotedblright} (i.e., talent recruiters of CERN students
and young researchers), collection and analysis of more than one
hundred documents (mostly internal CERN and Collaboration reports, but
also previous technical reports and research papers), and access to
different statistical databases, including the analysis of large
samples of company accounts data from the Orbis international dataset
(BvD). 

The evidence collected has then been structured in the form of a
computable model in matrix form, where each cell corresponds to a
benefit or cost variable and a year from 1993 to 2025, and beyond for
certain variables, such as human capital effects. Past missing data in
some years have been estimated and data for future years forecast by
simple models, as explained in detail below. 

While in general for past data and for minor items the baseline has been
taken as deterministic, for the critical forecasts a probability
distribution function (PDF) has been assigned, based either on the
sample information or on expert assessment of possible ranges of values
around the baseline. In practice, only some critical variables need to
be treated as stochastic (European Commission 2014). For a total of 19
variables (as reported below in the relevant section) a PDF has been
assumed based on expert data evaluation. To simplify computation, we
often assumed that a normal distribution is adequately proxied by a
triangular PDF\footnote{ While we mostly use triangular PDF for
computational reasons, we have checked that our overall results are
robust if we use other distributions.} (maximum, minimum, and mode
value, not always with symmetric tails), but in other cases we
considered that different distributions were more appropriate. We have
tested that in general using (truncated) normal distributions or other
continuous PDFs within a fixed range would not significantly change our
results. Finally, we have determined the PDF for the $NPV$ as for Eq.~(\ref{eq:npvres})
by running a Monte Carlo simulation (10,000 draws conditional on the
stochastic variables). 

Monte Carlo methods are the standard approach for the CBA of
infrastructure (Eckhardt 1987, Pouliquen 1970, Salling and Leleur 2011,
European Commission 2014, Florio 2014) as they allow estimating the
expected value of the variable of interest, with the overall
forecasting  accuracy conditional on the residual error of the assumed
PDF of the input variables and with  Monte Carlo error that has limit
zero for infinite draws. As some decision makers (European Commission
2014) are accustomed to consider performance variables in the form of
the internal rate of return ($IRR$, i.e., the value of $r$ such
that $NPV=0$) or the benefit/cost ratio, we have run Monte Carlo
simulations on these variables as well. Thus, we are able to generate a
conditional PDF of the $NPV$, the $IRR$, and the $B/C$ ratio\footnote{
Details of the MC simulations are available from the authors for all
the variables.}. 

In the next sections, for each contribution on the r.h.s. of Eq.~(\ref{eq:npvres}) we
present our estimation of the corresponding present value (PV).

\section{Costs}

LHC costs include past and future capital and operational expenditures
born by CERN and the Collaborations for building, upgrading, and
operating the machine and conducting experiments, including in-kind
contributions, for which there exists no integrated accounting. Three
categories of costs have been considered: i) construction capital
costs, ii) upgrade capital costs\footnote{ Only upgrade costs related
to the so-called {\textquotedblleft}Phase 1{\textquotedblright} have
been considered, these being sustained to optimize the physics
potential of LHC experiments for operation at higher luminosity.}, and
iii) operating costs. We have estimated CERN costs from the start
(1993) up to 2025, while for the different Collaborations (having
different reporting systems) we have reconstructed costs using their
own financial reports, supplemented by our assumptions for years after
2013. Integrated past flows have been capitalized to  ${t}_{0}$ = 2013
with a 0.03 social discount rate (in line with European Commission,
2014). Future costs have been discounted to 2013 euro values by a 0.03
social discount rate as well. 

In detail, we have estimated capital and operational expenditures
related to LHC as follows. Budgetary allocations from CERN to LHC have
been recovered from data communicated to us by the CERN Resource
Planning Department, drawing from the CERN Expenditure Tracking (CET)
system (account category, type, year, program at 31 March 2014). These
data cover all CERN program and subprogram expenditures in current CHF,
from January 1993 to 31 December 2013. The programs include:
Accelerators, Administration, Central Expenses, Infrastructure,
Outreach, Pension Fund, Research, and Services. Cost for each program
is disaggregated in various subprograms (e.g., under Accelerators there
are 19 subprograms, such as the SPS Complex, LHC, LEP, General R\&D,
etc.). In turn, each of these items shows expenditures on materials,
personnel, financial costs, and others, broken down into recurrent and
non-recurrent expenditure. We have excluded financial costs (such as
bank charges and interests) for the reasons explained in the previous
section; we have then identified the expenditure that can be attributed
to the LHC, rather than other CERN activities. 

In many cases, it was necessary to estimate an apportionment share to
the LHC of the expenditure for each item. To double-check these
accounting data, we have interviewed CERN staff in different
departments to ask whether the internal reporting actually covered all
the expenditures attributable to the LHC. The results of this data
collection process are provided in Table 1\footnote{ Current CHF values
have first been accounted in constant 2013 CHF by considering the
yearly change of average consumption prices from IMF World Economic
Outlook (October 2013), then expressed in euros at the exchange rate
$1~{\rm CHF}=0.812$~\euro\ (European Central Bank, average of daily rates for year
2013:
http://www.ecb.europa.eu/stats/exchange/eurofxref/html/eurofxref-graph-chf.en.html).}.
Some overheads are not recorded in internal reporting as being related
to specific accelerators or programs. However, as the LHC to some
extent increased CERN{\textquoteright}s administrative costs, given the
observation of past trends before the start-up of LHC operations, we
have attributed 10\% of CERN Administration, Central expenses,
Administrative and Technical personnel to the LHC. A sensitivity
analysis of the impact of apportioning a higher share of overhead costs
to LHC shows that the $NPV$ remains positive up to a 75\% share
attributed to LHC, without changing any other hypothesis. We have
identified scientific personnel costs of CERN from the reports of CERN
Personnel Statistics, available for each year. The share of this part
of the personnel every year is between 19\% in 1993 and 32\% in 2013.
This share of costs is assumed to balance with the contribution of CERN
scientists to the direct value of the LHC publications, similarly to
what we assume for non-CERN scientists in the Collaborations. We
discuss this assumption in detail in the next section, where we discuss
the valuation of scientific output (publications and other forms).

To these direct CERN costs we have added past in-kind contributions from
member and non-member states\footnote{ These are mainly in the form of
equipment made available for free to CERN by third parties and for
which in Annual Accounts (Financial Statements) 2008 (CERN/2840
CERN/FC/5337) a cumulative asset value of 1.47 MCHF is recorded,
combining in kind-contribution to the LHC machine and the detectors.
The attribution year by year of this cumulative figure has been done
assuming the same trend as for CERN procurement expenditures. }. We
have not included any forecasts of further in-kind contributions in
future. The forecast for 2014-2025 of CERN expenditures has been
communicated to us by CERN staff\footnote{ Based on the Draft Medium
Term Plan 2014 (personal communication April 2 2014). Again, we have
implemented an apportionment to LHC of each expenditure item. As all
values were given to us in constant CHF 2014, these were first
converted to CHF 2013 and then future values discounted to 2013 levels
by the 0.03 rate.}.

For the expenditures of the Collaborations, we have focused the analysis
on the four main experiments (ATLAS, CMS, ALICE, LHCb), as the
remaining ones (LHCf, FELIX, FP420, HV-QF, MOEDAL, TOTEM) are
comparatively quite small in terms of capital and operating costs. The
benefits of these experiments are also excluded from the computation of
the $NPV$. Our sources for the main four experiments have been the
Resource Coordinators of each
Collaboration\footnote{\textstylefootnotereference{ }We have analysed
the expenditure data particularly from these sources: CMS Summary of
Expenditure for CMS Construction for the Period from 1995 to 2008
(CERN-RBB-2009-032); CMS upgrade status report (CERN-RBB-2014-056);
Draft Budget for CMS Maintenance \& Operations in the Year 2014
(CERN-RBB-2013-086); Addendum No. 6 to the Memorandum of Understanding
for Collaboration in the Construction of the CMS Detector
(CERN-RBB-2013-070/REV); Addendum No. 7 to the Memorandum of
Understanding for Collaboration in the Upgrade of the CMS Detector
(CERN-RBB-2013-127); Addendum No. 8 to the Memorandum of Understanding
for Collaboration in the Upgrade of the CMS Detector
(CERN-RBB-2013-128); Memorandum of Understanding for Maintenance and
Operation of the ATLAS Detector (CERN-RBB-2002-035); ATLAS Upgrade
Status Report 2013-2014 (CERN-RBB-2014-022); Request for 2014 ATLAS
M\&O Budget (CERN-RBB-2013-079); Memorandum of Understanding for
Maintenance and Operation of the LHCb Detector
(CERN-RRB-2002-032.rev-2008); Addendum No. 01 to the Memorandum of
Understanding for the Collaboration in the Construction of the LHCb
detector (CERN/RBB 2012-119A.rev-2014); Status of the LHCb upgrade
(CERN-RRB-2014-033); RRB Apr.2014 (CERN-RRB-2014-039); for ALICE data,
the source is a personal communication (May 7 2014) comprising data
such as Core Expenditure 2007-2013, Construction costs, including
Common Fund, per system, M\&O A-budget and B-budget. Fifteen more
reports have been processed by us for the analysis of costs (a detailed
list is available from the authors upon request).}. Forecasts of future
expenditures of the Collaborations have been based on the same sources.
When only cumulative data at a certain year were available, the yearly
distribution has been interpolated linearly. In the same way, some
missing yearly data for the LHCb Collaboration have been interpolated. 

We have not considered the cost implications of new projects---the High
Luminosity Project and of the LHC Upgrade Phase 2---as they mostly will
run after our time horizon. To avoid double counting, the CERN
contributions to the Collaborations have been excluded from their total
expenditures. Similarly to what we assume for CERN, the scientific
personnel cost of the Collaborations (paid by their respective
institutes) has been taken as balancing the marginal cost valuation of
the scientific publications attributed to each experiment and excluded
from the grand total of cost; see the next section regarding this
point. 

\begin{figure}
\includegraphics[width=6.1689in,height=3.1228in]{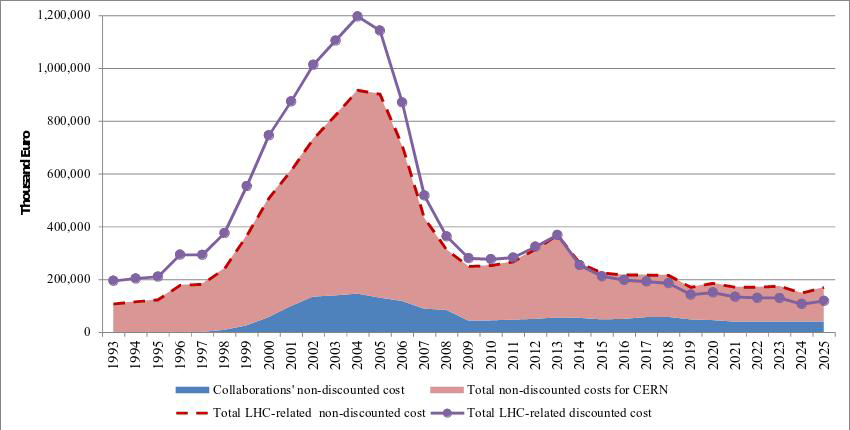}

\caption{ Time distribution of LHC costs (discounted and
non-discounted).} 
\label{fig:costs}
\end{figure}

The overall trend of cumulated LHC-related CERN and Collaboration
expenditures is shown in Fig.~\ref{fig:costs}. While we consider the information up
to 2013 as given, we treat the forecasts from 2014-2025 as stochastic.
We have assumed a normal distribution of the total future cost
(2014-2025) with mean equal to\footnote{Here and in the sequel we will
  use the notation G\euro, M\euro, k\euro for billion, million and
  thousands of euro.}1.97 G\euro\  and a standard
deviation compatible with mean {\textpm}50\% as asymptotic values. This
range is based on in-depth interviews with experts at CERN and analysis
on the most optimistic and pessimistic future cost scenarios\footnote{
As mentioned in the previous section, we have not included
decommissioning costs as we have no reliable information on them. For
the same reason, we have also not tried to forecast accidents or
negative externalities.}. 

Summing up: after including the value of in-kind contributions, we have
reconstructed the time distribution of LHC costs over 1995 to 2008 (see
Fig.~\ref{fig:costs}), while CERN costs unrelated to LHC and costs for future
upgrades have been excluded, as their benefits will occur beyond our
time horizon. Our final estimate for the expected mean value of the
total cost of the LHC over 33 years (1993-2025) is  $\left\langle
K+LO+O\right\rangle = 13.5~G$~\euro,
net of scientific personnel cost\footnote{ To put this figure in
perspective, it would be interesting to compare it with other
large-scale scientific programs. This comparison is beyond our research
scope, but just to mention one figure, the yearly budget of NASA
(comprising several programs) in 2014 was USD 17.6~M\$
(http://ww.nasa.gov/sites/default/files/files/NASA\_2016\_Budget\_Estimates.pdf.
Thus the total cost of the LHC over more than 30 years is roughly of
the same size as one year of the NASA budget.}.\textbf{ }Here and in
the next sections, the mean value refers always to the outcome of the
Monte Carlo process after 10,000 draws.

\begin{table}{\tiny
\begin{center}
\tablehead{}
\begin{supertabular}{|m{1.8511599in}m{1.2094599in}m{0.9483598in}m{1.0031599in}m{0.79275984in}|}
\hline
\multicolumn{1}{|m{1.8511599in}|}{\raggedleft
\textbf{\textcolor{black}{A}}\textbf{\textcolor{black}{pp}}\textbf{\textcolor{black}{o}}\textbf{\textcolor{black}{rtionment}}\textbf{\textcolor{black}{
}}\textbf{\textcolor{black}{s}}\textbf{\textcolor{black}{hare}}} &
\multicolumn{1}{m{1.2094599in}|}{\raggedleft \textbf{\textcolor{black}{
Apportionment share}}} &
\multicolumn{1}{m{0.9483598in}|}{\raggedleft
\textbf{\textcolor{black}{LHC-related non-recurrent expense}}\par

\raggedleft \textbf{\textcolor{black}{expense}}} &
\multicolumn{1}{m{1.0031599in}|}{\raggedleft
\textbf{\textcolor{black}{LHC-related recurrent expense}}} &
\raggedleft\arraybslash
\textbf{\textcolor{black}{L}}\textbf{\textcolor{black}{H}}\textbf{\textcolor{black}{C}}\textbf{\textcolor{black}{{}-related
total}}\textbf{\textcolor{black}{
}}\textbf{\textcolor{black}{cost}}\\\hline
\multicolumn{1}{|m{1.8511599in}|}{\textbf{Accelerators}} &
\multicolumn{1}{m{1.2094599in}|}{} &
\multicolumn{1}{m{0.9483598in}|}{\raggedleft \textbf{4,486,682}} &
\multicolumn{1}{m{1.0031599in}|}{\raggedleft \textbf{1,690,053}} &
\raggedleft\arraybslash
\textbf{6,}\textbf{1}\textbf{76,}\textbf{7}\textbf{36}\\\hline
CLIC &
\raggedleft 0\% &
\raggedleft 0 &
\raggedleft 0 &
\raggedleft\arraybslash 0\\
CNGS &
\raggedleft 0\% &
\raggedleft 0 &
\raggedleft 0 &
\raggedleft\arraybslash 0\\
Consolidation &
\raggedleft 100\% &
\raggedleft 146,370 &
\raggedleft 630 &
\raggedleft\arraybslash 146,999\\
Experimental Areas PS &
\raggedleft 0\% &
\raggedleft 0 &
\raggedleft 0 &
\raggedleft\arraybslash 0\\
Experimental Areas SPS &
\raggedleft  ${0\mathit{}}^{a}$ and  ${50}^{b}$ &
\raggedleft 2,664 &
\raggedleft 50,911 &
\raggedleft\arraybslash 53,575\\
General R\&D &
\raggedleft 0\% {\textless}2007; 50\% from 2008 &
\raggedleft 1,760 &
\raggedleft 727 &
\raggedleft\arraybslash 2,487\\
General Services &
\raggedleft 0\% {\textless}2007; 50\% from 2008 &
\raggedleft 1,480 &
\raggedleft 11,052 &
\raggedleft\arraybslash 12,533\\
LEP &
\raggedleft 0\% &
\raggedleft 0 &
\raggedleft 0 &
\raggedleft\arraybslash 0\\
LHC &
\raggedleft 100\% &
\raggedleft 4,076,429 &
\raggedleft 1,111,295 &
\raggedleft\arraybslash 5,187,724\\
LHC injectors &
\raggedleft 100\% &
\raggedleft 28,420 &
\raggedleft 3,221 &
\raggedleft\arraybslash 31,641\\
LHC injectors upgrade &
\raggedleft 100\% &
\raggedleft 14,103 &
\raggedleft 186 &
\raggedleft\arraybslash 14,289\\
LHC upgrade &
\raggedleft 100\% &
\raggedleft 153,252 &
\raggedleft 3,218 &
\raggedleft\arraybslash 156,470\\
Low and medium energy &
\raggedleft 0\% &
\raggedleft 0 &
\raggedleft 0 &
\raggedleft\arraybslash 0\\
Medical applications &
\raggedleft 0\% &
\raggedleft 0 &
\raggedleft 0 &
\raggedleft\arraybslash 0\\
PS complex &
\raggedleft 50\% &
\raggedleft 25,242 &
\raggedleft 231,207 &
\raggedleft\arraybslash 256,449\\
R\&D &
\raggedleft 50\% &
\raggedleft 2,944 &
\raggedleft 2,797 &
\raggedleft\arraybslash 5,741\\
R\&D CLIC &
\raggedleft 0\% &
\raggedleft 0 &
\raggedleft 0 &
\raggedleft\arraybslash 0\\
SPS complex &
\raggedleft  ${50}^{c}$ and  ${80}^{d}$  &
\raggedleft 34,020 &
\raggedleft 274,809\par

 &
\raggedleft\arraybslash 308,829\\\hline
\textbf{A}\textbf{dm}\textbf{inistration} &
 &
\raggedleft \textbf{9,325} &
\raggedleft \textbf{314,484} &
\raggedleft\arraybslash \textbf{323,809}\\\hline
Administrative computing &
\raggedleft 25\% &
\raggedleft 1,855 &
\raggedleft 36,585 &
\raggedleft\arraybslash 38,440\\
Directorate &
\raggedleft 25\% &
\raggedleft 3,438 &
\raggedleft 84,329 &
\raggedleft\arraybslash 87,767\\
Finances &
\raggedleft 25\% &
\raggedleft 716 &
\raggedleft 30,729 &
\raggedleft\arraybslash 31,444\\
General Services &
\raggedleft 25\% &
\raggedleft 1,400 &
\raggedleft 24,705 &
\raggedleft\arraybslash 26,105\\
HR &
\raggedleft 25\% &
\raggedleft 1,801 &
\raggedleft 113,267 &
\raggedleft\arraybslash 115,068\\
Procurement &
\raggedleft 25\% &
\raggedleft 115 &
\raggedleft 24,869 &
\raggedleft\arraybslash 24,984\\\hline
\textbf{Cent}\textbf{r}\textbf{al}\textbf{
}\textbf{ex}\textbf{p}\textbf{enses} &
 &
\raggedleft \textbf{268} &
\raggedleft \textbf{91,559} &
\raggedleft\arraybslash \textbf{91,827}\\\hline
bank charges and interests &
\raggedleft 0\% &
\raggedleft 0 &
\raggedleft 0 &
\raggedleft\arraybslash 0\\
Centralised personnel Expenses &
\raggedleft 25\% &
\raggedleft 0 &
\raggedleft 56,968 &
\raggedleft\arraybslash 56,968\\
Housing fund &
\raggedleft 0\% &
\raggedleft 0 &
\raggedleft 0 &
\raggedleft\arraybslash 0\\
Insurances &
\raggedleft 25\% &
\raggedleft 0 &
\raggedleft 14,111 &
\raggedleft\arraybslash 14,111\\
Internal taxation &
\raggedleft 0\% &
\raggedleft 0 &
\raggedleft 0 &
\raggedleft\arraybslash 0\\
phone and postal charges &
\raggedleft 25\% &
\raggedleft 0 &
\raggedleft 1,101 &
\raggedleft\arraybslash 1,101\\
Storage management &
\raggedleft 25\% &
\raggedleft 268 &
\raggedleft 19,379 &
\raggedleft\arraybslash 19,647\\\hline
\textbf{I}\textbf{nfrastr}\textbf{u}\textbf{ctu}\textbf{r}\textbf{e } &
 &
\raggedleft \textbf{181,721} &
\raggedleft \textbf{1,092,689} &
\raggedleft\arraybslash \textbf{1,274,410}\\\hline
Building construction &
\raggedleft 80\% &
\raggedleft 69,728 &
\raggedleft 0 &
\raggedleft\arraybslash 69,728\\
Computing &
\raggedleft 20\% &
\raggedleft 5,124 &
\raggedleft 27,702 &
\raggedleft\arraybslash 32,826\\
Energy &
\raggedleft 20\%{\textless}2000, then 50\%, 80\% as of 2008 &
\raggedleft 155 &
\raggedleft 478,824 &
\raggedleft\arraybslash 478,979\\
General Services &
\raggedleft 50\% &
\raggedleft 0 &
\raggedleft 438 &
\raggedleft\arraybslash 438\\
Medical service &
\raggedleft 20\%{\textless}2000, then 50\%, 80\% as of 2008 &
\raggedleft 6,497 &
\raggedleft 108,786 &
\raggedleft\arraybslash 115,284\\
Site facility &
\raggedleft 40\% &
\raggedleft 83,850 &
\raggedleft 468,111 &
\raggedleft\arraybslash 551,961\\
Technical infrastructure &
\raggedleft 40\% &
\raggedleft 10,144 &
\raggedleft 0 &
\raggedleft\arraybslash 10,144\\
Waste management &
\raggedleft 40\% &
\raggedleft 6,223 &
\raggedleft 8,828 &
\raggedleft\arraybslash 15,050\\\hline
\textbf{O}\textbf{ut}\textbf{r}\textbf{each } &
 &
\raggedleft \textbf{20,053} &
\raggedleft \textbf{141,812} &
\raggedleft\arraybslash \textbf{161,865}\\\hline
Communication &
\raggedleft 80\% &
\raggedleft 15,274 &
\raggedleft 104,498 &
\raggedleft\arraybslash 119,772\\
Exchanges &
\raggedleft \ \ \ \ 0\% &
\raggedleft 0 &
\raggedleft 0 &
\raggedleft\arraybslash 0\\
Knowledge and Technology Transfer &
\raggedleft \ \ \ \ 50\% &
\raggedleft 4,779 &
\raggedleft 18,306 &
\multicolumn{1}{m{0.79275984in}}{\raggedleft\arraybslash 23,085}\\
Schools &
\raggedleft 0\% &
\raggedleft 0 &
\raggedleft 0 &
\raggedleft\arraybslash 0\\\hline
\textbf{Pe}\textbf{n}\textbf{sion}\textbf{
}\textbf{F}\textbf{u}\textbf{n}\textbf{d} &
 &
\raggedleft \textbf{0} &
\raggedleft \textbf{0} &
\raggedleft\arraybslash \textbf{0}\\\hline
Pension fund &
\raggedleft 0\% &
\raggedleft 0 &
\raggedleft 0 &
\raggedleft\arraybslash 0\\\hline
\textbf{Res}\textbf{e}\textbf{a}\textbf{r}\textbf{ch} &
 &
\raggedleft \textbf{618,001} &
\raggedleft \textbf{2,533,356} &
\raggedleft\arraybslash \textbf{3,151,357}\\\hline
Computing &
\raggedleft  ${50}^{e}$ and  ${80}^{f}$ &
\raggedleft 23,854 &
\raggedleft 71,736 &
\raggedleft\arraybslash 257,658\\
Controls &
\raggedleft 80\% &
\raggedleft 26 &
\raggedleft 71,736 &
\raggedleft\arraybslash 3,385\\
Data analysis &
\raggedleft  ${0}^{g}$,  ${50}^{h}$, ${80}^{i}$ and  ${100}^{l}$ &
\raggedleft 8,959 &
 &
\raggedleft\arraybslash 80,695\\
Electronics &
\raggedleft 50\% &
\raggedleft 5,498 &
\raggedleft 1,192 &
\raggedleft\arraybslash 148,102\\
EU supported R\&D general &
\raggedleft 50\% &
\raggedleft 25,572 &
\raggedleft 291,565 &
\raggedleft\arraybslash 26,763\\
General Services\ \  &
\raggedleft 50\% &
\raggedleft 26,345 &
\raggedleft 2,813 &
\raggedleft\arraybslash 317,910\\
Grid computing &
\raggedleft 80\% &
\raggedleft 1,447 &
\raggedleft 161,380 &
\raggedleft\arraybslash 4,260\\
LHC computing &
\raggedleft 100\% &
\raggedleft 126,539 &
\raggedleft 1,252,968 &
\raggedleft\arraybslash 287,919\\
LHC detectors &
\raggedleft 100\% &
\raggedleft 317,039 &
\raggedleft 272,638 &
\raggedleft\arraybslash 1,570,007\\
LHC detectors upgrade &
\raggedleft 100\% &
\raggedleft 78,328 &
\raggedleft 0 &
\raggedleft\arraybslash 350,966\\
Non-LHC physics &
\raggedleft 0\% &
\raggedleft 0 &
\raggedleft 99,297 &
\raggedleft\arraybslash 0\\
Theoretical physics &
\raggedleft 50\% &
\raggedleft 4,394 &
\raggedleft 17,441 &
\raggedleft\arraybslash 103,691\\\hline
\textbf{Servic}\textbf{e}\textbf{s} &
 &
\raggedleft \textbf{3,039} &
\raggedleft \textbf{17,441} &
\raggedleft\arraybslash \textbf{20,480}\\\hline
Electronics &
\raggedleft 80\% &
\raggedleft 3,039 &
\raggedleft 17,441 &
\raggedleft\arraybslash 20,480\\\hline
\textbf{\textcolor{black}{ Total}} &
 &
\raggedleft \textcolor{black}{5,319,088} &
\raggedleft \textcolor{black}{5,881,396} &
\raggedleft\arraybslash \textcolor{black}{11,200,484}\\\hline
\end{supertabular}
\end{center}}
\caption{
LHC-related costs covered by CERN by Programme and Subprogrammes and
apportionment share to LHC (1003-2013; k\euro\ at 2013 constant
prices).
Source: Author's elaboration based on CERN data and
interviews, see main text. \textit{a}= Codes EP, EPL, EPP. \textit{b}=
Codes ASE, ATB ESI. \textit{c=}Codes FSP, RFT. \textit{d=} Codes ASM
FAS, RFS, TSP. \textit{e=} Codes RSC, RSI. \textit{f}= Codes RCE, RCG,
RCL. \textit{g}= Code RCX. \textit{h}= Code RRD. \textit{i}= Codes RDD,
RDH.
}\end{table}

\section{Benefits to scientists: the value of publishing }

We start our discussion of benefits with one that turned out to be
small, when properly measured: the benefit of academic publishing per
se. In fact, the core benefit of the LHC to scientists is the
generation of experimental data that sustain the opportunity to publish
new research. It is important to clarify that we are not valuing here
the wider social impact of the actual content of the publications,
i.e., of the scientific value per se or of its future practical use (if
any), but we focus only on the direct effect of publications for
science insiders, a special social group. 

We briefly elaborate on this issue. The paper by Peter Higgs,
introducing in a short paragraph the theory of a massive boson, was
published in 1964, about the same time as papers by other physicists
now acknowledged as leading to a similar theoretical prediction. It
took nearly 50 years to confirm this intuition experimentally at the
LHC. Nobody currently knows if and when the theoretical prediction of a
new particle decades ago, its recent experimental discovery, and
further precision measurements in future, will lead to any practical
application. We know, for example, that more than one hundred years
after the pathbreaking articles by Alfred Einstein (1905 and 1916),
practical applications of the theories of special and general
relativity, respectively, are now widespread, e.g., in any GPS device.
However, this ex-post (after one hundred years) knowledge is not
helpful to evaluate ex-ante the social impact of a specific
publication, or of any number of publications: it only suggests that
there is a non-zero chance that any substantial new knowledge will have
an economic impact, which seems a reasonable assumption. 

Instead, for the scientists, either CERN employees or those hired by the
universities and other institutes participating in the experiments, the
direct benefit of publications in principle is measurable by the track
record of past publications. In this perspective, the benefit of
publications is proxied by its impact on the scientific community. 
This benefit has a limited impact on the overall balance of the social
impact of the LHC. This is not surprising because, first, the
scientific community of high-energy physicists and related fields is
small relative to other social groups; and because we are not including
here the value of knowledge to society per se as embodied in the
publication (see our discussion of non-use benefits for the taxpayers).

The observable demand to publish and to access scientific publications
does not provide a set of market prices that can be used to estimate
the marginal willingness to pay by science insiders. For example, the
subscription prices of journals are usually paid by libraries for their
users, the open source fee for some journals may be paid by research
funds, many papers are available for free (e.g., the more than one
million pre-prints available in the ArXiv repository for physics). The
usual alternative to estimating WTP by revealed or stated preferences
is the marginal cost approach to the estimation of benefits (European
Commission 2014). Hence, a publication produced by LHC insider
scientists ( ${L}_{0}$) has a value that is on average equal (or not
less) to its production costs (scientific personnel costs). In other
words, the marginal social value of a
{\textquotedblleft}statistical{\textquotedblright} publication is the
average marginal cost of producing such a publication. Assuming
linearity of publication production respect to time (which is well
documented by the stability of the average coefficient of number of
publications per researcher per year in each field---see Carrazza et
al. 2014), this fact has the interesting consequence that the
scientific personnel cost is balanced by the benefit of the
publications. Thus, with a considerable advantage in terms of
estimation of costs and benefits, the two amounts can be assumed to
cancel out; and therefore neither the benefits nor the costs are
explicitly included in the CBA. While some evaluator may try separately
to estimate the cost and benefits (in our narrow meaning) of the
publications (and similar products, such as preprints, conference
abstracts, etc.) and further refine the analysis, this estimation in
the LHC case would be an overwhelming task, given the large numbers of
scientists involved in the experiments at any point of time (around
10,000). 

\begin{figure}
\includegraphics[width=5.1346in,height=1.9in]{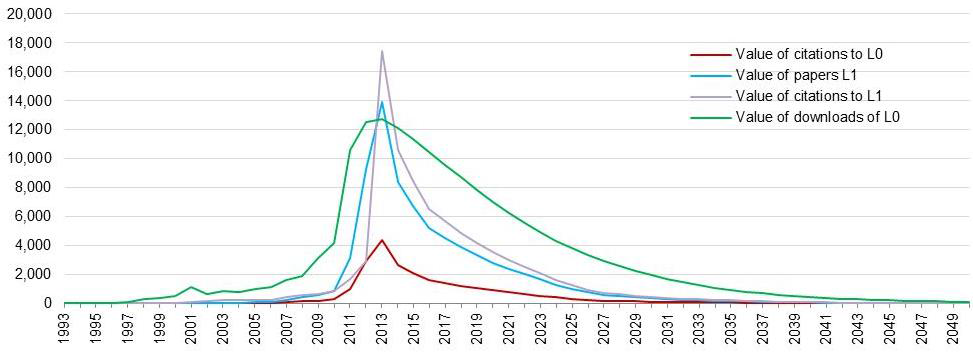}

\caption{Economic value (constant k\euro\ 2013) per year of
citations to  ${L}_{0}$ and  ${L}_{1}$ papers; value of  ${L}_{1}$
papers; value of downloads of ${L}_{0}$ papers.}
\label{fig:papers}
\end{figure}

Hence, we exclude from the benefits the first round of publications
\textit{L}\textit{\textsubscript{0}}, those from the LHC insiders, and
consider then only the additional benefits arising from papers (
${L}_{1}$) by non-LHC scientists citing  ${L}_{0}$ papers, with the
direct benefit of further papers (\textit{L}\textit{\textsubscript{2}})
citing  ${L}_{1}$ papers in turn considered to be negligible (to be
conservative), but including their citations to
\textit{L}\textit{\textsubscript{1}} papers. We proxy the MSV of 
${L}_{1}$ papers through the average salary received by an average
scientist for the time spent on doing research and writing a paper. Our
forecast of outputs is based on an estimate of publication trajectories
obtained through a statistical model over a period of \textit{N} = 50
years, starting in 2006. The results are summarized in Fig.~\ref{fig:papers}. We
explain below the procedure in detail.

The past (1993-2012) number of LHC-related scientific publications 
${L}_{0}$ (including CERN and Collaborations) has been extracted from
the inSPIRE database (http://inspirehep.net/) by Carrazza et al. 2014.
The data include both published articles and preprints. Citations of
these up to 2012 have been retrieved from the same source. In order to
forecast the number ${EL}_{0}$ of  ${L}_{0}$ publications 2013-2025,
we have applied a double exponential model (Bacchiocchi and Montobbio
2009, Carrazza et al. 2014). This model is based on a calibration of
the publishing trajectory of the LHC predecessor at CERN, the LEP
accelerator. It takes the form:
\begin{equation}
\label{eq:biblio}
EL_0(t)=\alpha_1\alpha_2 e^{-\beta_1(T-t)}\left[1-
e^{-\beta_2(T-t)}\right],
\end{equation}
\begin{itemize}
\item  $\alpha_1= 65000$ is the expected total number of authors
of publications during the entire time span considered;
\item  $\alpha_2= 2$ is a proxy of their productivity; 
\item  $\beta_1= 0.18$ and  $\beta_2= 0.008$ are two
parameters determining the shape of the curve, based on the observed
pattern of publications related to the LEP;
\item  $T = 50$ is the total number of years;
\item  $t=\left(0,{\dots},50\right)$ is the number of remaining years
from 2006, the start year of estimations, to the end of the simulation
period (2056). 
\end{itemize}
All the parameters are estimated from the data, except  ${\beta }_{1}$
and  ${\beta }_{2}$ .

The forecast of the number of  ${L}_{1}$ publications over the years
2013-2050 has been based on observed pattern of average number of
citations per paper (inSPIRE data), without assuming any new spike
after the one related to the discovery of Higgs boson. This is again a
conservative assumption, because it amounts to saying that nothing of
importance will be discovered by the LHC until 2025. 

We have then estimated the citations to  ${L}_{1}$ papers by  ${L}_{2}$
papers. Again, the number of  ${L}_{2}$ papers until 2012 is based on
inSPIRE, while to forecast 2013-2050 we assume 4 citations per paper,
in line with the previous years. To these figures we have added total
arXiv downloads for the field of High Energy Physics\footnote{ Data
have been provided to us by Cornell University Library upon request.},
which we used for 1994-2013, while in order to forecast until 2050 we
have assumed the same average in future as the past (64 downloads per
paper). This average number of downloads has been applied to  ${L}_{0}$
papers. 

To sum up, the direct benefits for insiders of the science community are
thus: the value of  ${L}_{1}$ papers; the value of  ${L}_{1}$ citations
and downloads to  ${L}_{0}$ papers; and the value of  ${L}_{2}$
citations to  ${L}_{1}$ papers. As mentioned, the value of  ${L}_{0}$
papers cancels out their production cost and it is not included. The
value of  ${L}_{2}$ papers and beyond, and citations to them, is
considered to be negligible. All values are discounted at the 3\%
social discount rate. 

After the baseline estimations, risk analysis has been performed on the
total PV of the publications. To perform a Monte Carlo simulation, a
PDF for the following variables has been assumed: number of references
to  ${L}_{0}$ papers in papers  ${L}_{1}$; percentage of time of
scientists devoted to research papers produced per year per average
salary of non-LHC scientists; time per download time per citation.

The resulting total present value of the publications has a mean
277~M\euro. The value of 
publications (net of \textit{L}\textit{\textsubscript{0}}) per se pays
back only a tiny fraction of around 2\% of the total cost (net of
scientific personnel costs).

\section{Benefits to students and post-docs: human capital
formation }

\begin{figure}

\includegraphics[width=4.7543in,height=3.3728in]{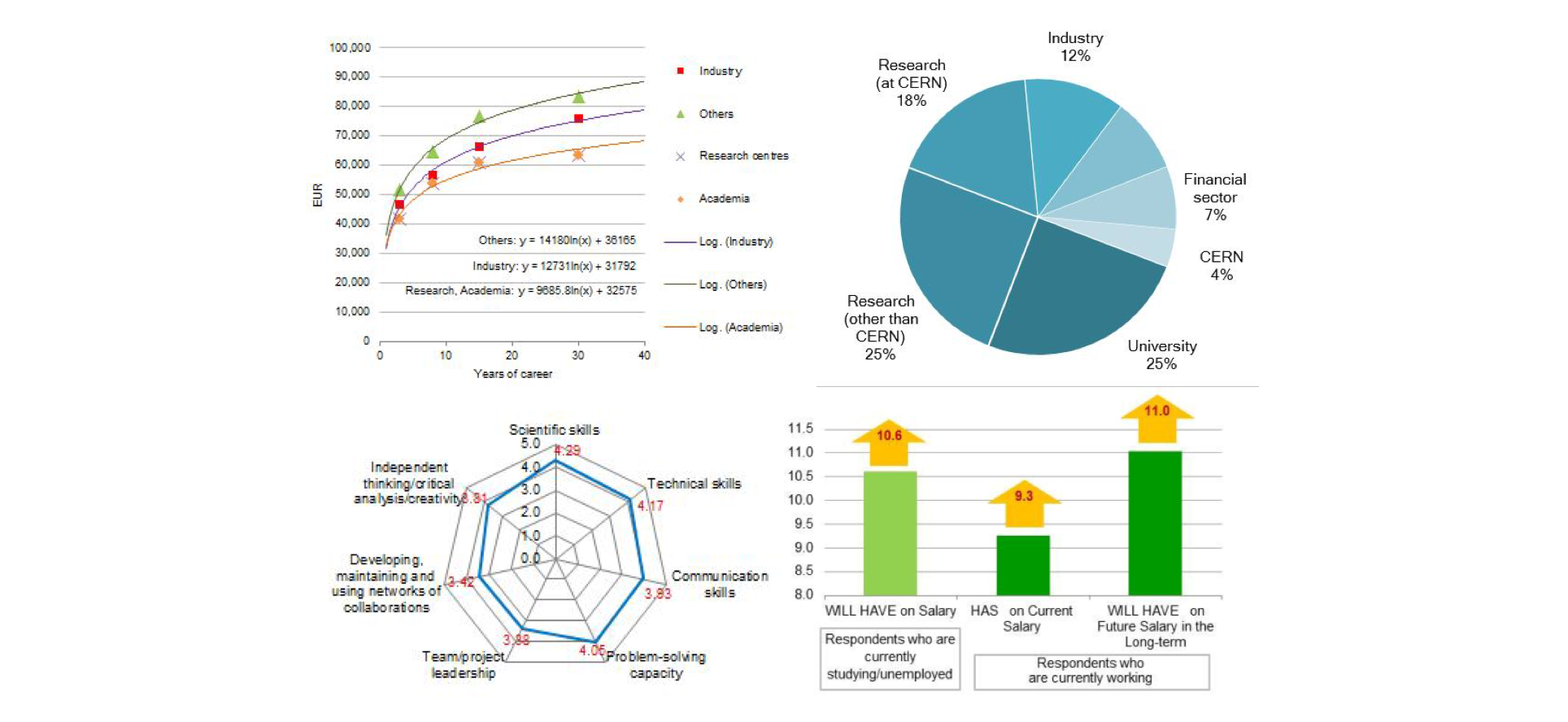}
\caption{Top: types and number of people benefitting from
training at the LHC, historical data and forecasts. Centre: estimation
of future average salaries (left); current employment sector of CERN
alumni (right). Bottom: perception of skill improvements due to the LHC
experience (left); percentage impact on salary due to the LHC
experience estimated by current students (light green) and
past-students (dark green) (right).
}\label{fig:hr}
\end{figure}

We have estimated that beneficiaries of human capital formation
(Schopper 2009, Camporesi 2001) at the LHC over the time period
1993-2025 include nearly 36,800 early-stage researchers (ESR): around
19,400 students and 17,400 post-docs (not including participants in
summer schools or short courses). Consistent with the literature on
marginal returns to education (see, e.g., Harmon 2011) the benefit
arising from ESR experience at the LHC is valued as the present value
of the LHC-related incremental salary earned over the entire work
career (see Fig.~\ref{fig:hr}). This effect obviously is not the full future
salary of former ESR, but it is an estimation of the LHC
{\textquotedblleft}premium{\textquotedblright} effect on future
earnings.

We have considered five types of ESR: CERN doctoral students; CERN
technical students; CERN fellows; users under 30 years; and users
between 30 and 35 years. The sources of data are the yearly reports of
CERN Personnel statistics from 1995 until 2013. We have estimated the
number of incoming students year by year for each type and average
stay, based on past data available at the CERN Human Resources
Department and from interviews with staff. Future incoming student
flows have been extrapolated from past trends and checked with CERN.
The HR Department records all types of students and post-docs, but we
need an apportionment of these flows to the LHC. We have computed such
apportionments with data from the Collaborations and additional
interviews at CERN, leading to the following estimates: 30\% of the
total flows (for the period 1993-1998); 50\% (1999-2001); 70\%
(2002-2007); and 85\% (2008-2025). The resulting figures have then been
attributed to each of the five types, based on the historical
distribution, in order to derive the flow of annual incoming students
over the years 1993-2025. 

 Ideally, in order to estimate the economic benefit to each of these
types of ESR, we would have needed a sample of former LHC students of
different cohorts in their present occupation and a control group of
non-LHC peers. However, given that most of the actual flows of incoming
ESR at the LHC occurred in recent years, i.e., after the startup of
experiments in 2008, the latter information is not available. Hence,
our strategy was to make an estimation based on two samples,
respectively of current and former students and post-docs. A survey,
directed at both students and former students, was performed between
May and October 2014 and in March 2015 through an on-line questionnaire
and direct interviews at CERN. The details of the survey, including the
questionnaire, are available in Catalano et al. (2015a). The survey
strategy was to elicit both expectations of current students at the LHC
and evaluations from former students, now employed in different jobs,
including outside academia. Information from 384 interviewees coming
from 52 different countries has been collected: 75\% of respondents are
male; 38\% are 20-29 years old, 43\% are 30-39 years old, the remaining
are more than 40 years old; 65\% of respondents are related to the CMS
Collaboration and 22\% to ATLAS, while the remainder are in other
experiments or LHC-related research at CERN. Each respondent has
answered questions on a number of individual characteristics, his/her
perception of the skills acquired at the LHC, and finally on an ex-ante
(students) or ex-post (former students) perceived LHC premium on their
salary. We assume that former ESR have some knowledge of job market
opportunities and can compare their expectations with those of their
peers. We have found that the two sample averages for the premium
effect are strikingly similar, as can be seen in Fig.~\ref{fig:hr}. This suggests
that information on job opportunities and salaries is widespread and
convergent (this is, after all, a relatively small international
network of young researchers with close formal and informal linkages).
Given that the former ESR at the LHC have gained actual experience of
the post-LHC career market, we have focused on the premium declared by
the respondents who have already found a job: the sample average is
equal to 9.3\% . 

This percentage premium has been applied to the average annual salary at
different experience levels, retrieved from the Payscale database. In
particular, we have classified salaries by experience level (entry,
mid-career, experienced, and late career) for different jobs in the
USA\footnote{ See, e.g.,
http://www.payscale.com/research/US/Job=Electronics Engineer/Salary)}
grouped in four broad sectors: industry, research centers, academia,
others (the latter including, for instance, finance, computing, and
civil service). A distribution of the number of CERN students across
these broad sectors has been retrieved based on earlier work by
Camporesi (2001) and other sources\footnote{ For CERN technical
students we have assumed that only 10\% will go either to research
centres or in academia, and 45\% respectively in the other two sectors;
for the other students, we have assumed a destination in research and
academia for 60\% and 20\% each for the others. Interviews with experts
(including {\textquotedblleft}head hunters{\textquotedblright} who
regularly monitor the CERN students) have confirmed this
distribution.}. The four aforementioned career points have been
interpolated with a logarithmic function. 

Given the average salary in each broad sector, the LHC premium declared
by interviewees, and the above-assumed shares of students finding a job
in each sector, we have computed a job effect component of the human
capital formation benefit. Considering that the difference between the
pay in research and academia and the two other sectors combined is
between 13\% and 18\% (increasing with the level of experience), and
that 14\% of the former students who have participated in the survey
have been diverted to better-paid jobs in industry or other sectors
(consistent with earlier findings by Camporesi 2001), an additional
small premium of between 2-3\% (triangular PDF with average and mode
both equal to 2.5\%) has been applied, because of the composition
effect across occupations. The resulting combined 11.8\% premium has
been attributed to an average student over a career spanning 40 years,
with the implication, for example, that the cohort of 2025 students
will enjoy the benefit up to 2065. Interestingly, this figure is well
in the range of the returns to higher education in the literature (for
a review of more than 50 years of empirical research, see Montenegro
and Patrinos 2014, who find the highest returns for tertiary
education). The total number of ESR, in turn, has been taken as a
triangular PDF with maximum and minimum equal to {\textpm}15\% of the
mode and mean, based on available data. All values are discounted,
which, because of the long time span, roughly halves the cumulative
benefit in comparison with its undiscounted
value.\textcolor[rgb]{0.14901961,0.14901961,0.14901961}{ }

The resulting mean value of the corresponding benefits is  $\left\langle
HC\right\rangle  = 5.5~G$~\euro. This social
benefit pays back 41\% of the total LHC social cost and this is the
largest component on the benefit side.

\section{Benefits to firms and other organizations: technological
spillovers}

There are two main types of beneficiaries of LHC on the business side:
firms in the procurement chain, because of learning-by-doing effects;
and other firms or professional organizations acquiring knowledge for
free. In both cases, these effects can be described as externalities
related to the transfer of knowledge to third parties outside or beyond
any contractual relations with the CERN.

Profits to firms directly arising from procurement to CERN are part of
the LHC costs and are not considered a social benefit (assuming that
there was no idle capacity in the firms). Thus, the benefits to
LHC-related supplier firms consist of incremental profits gained
through additional sales to third parties, after the procurement
contract with CERN, thanks to technology transfer and knowledge
acquired {\textquotedblleft}for free.{\textquotedblright} Such effects
become particularly important the more co-designed is the technology,
particularly because CERN almost never patents its own inventions, a
famous example being the World Wide Web (Schopper 2009, Boisot et al.
2011). We have briefly mentioned in section 2 the scope and scale of
technological innovation related to building and maintaining the LHC,
in collaboration with a large number of firms involved (more than 1500,
see below). 

\begin{figure}
\includegraphics[width=6.6929in,height=5.2346in]{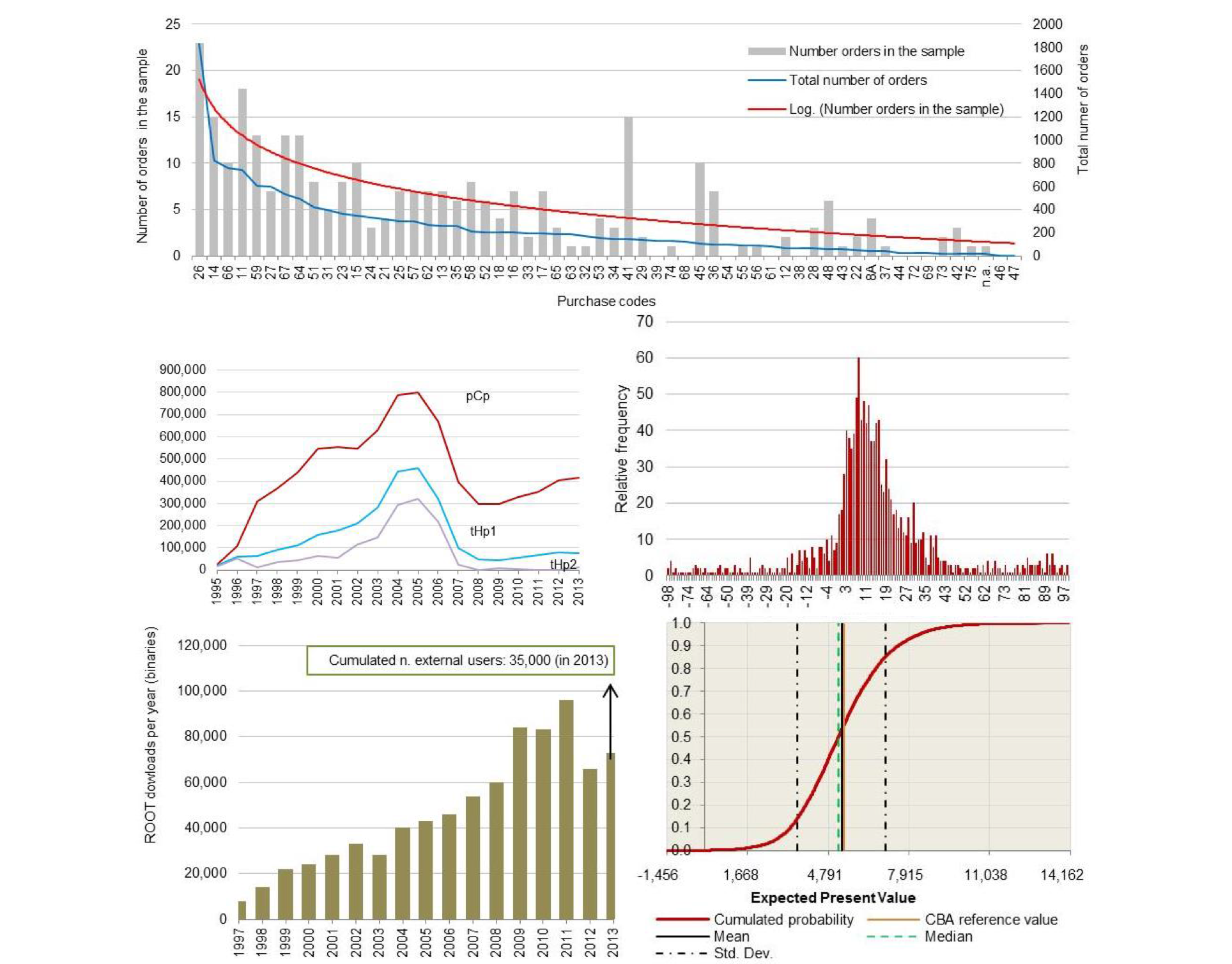}
\caption{Top: Benefits to firms in the CERN supply chain from a
sample of 300 orders by purchase code compared with all LHC orders
{\tiny (CERN activity codes: 11 building work - 12 roadworks - 13 installation
and supply of pipes - 14 electrical installation work - 15 heating and
air-conditioning equipment (supply and installation) - 16 hoisting gear
- 17 water supply and treatment - 18 civil engineering and buildings -
21 switch gear and switchboards - 22 power transformers - 23 power
cables and conductors - 24 control and communication cables - 25 power
supplies and converters - 26 magnets - 27 measurement and regulation -
28 electrical engineering - 29 electrical engineering components - 31
active electronic components - 32 passive electronic components - 33
electronic measuring instruments - 34 power supplies - transformers -
35 functional modules \& crates - 36 rf and microwave components and
equipment - 37 circuit boards - 38 electronics - 39 electronic assembly
and wiring work - 41 computers and work-stations - 42 storage systems -
43 data-processing peripherals - 44 interfaces (see also 35 series) -
45 software - 46 consumables items for data-processing - 47 storage
furniture (data-processing) - 48 data communication - 51 raw materials
(supplies) - 52 machine tools, workshop and quality control equipment -
53 casting and moulding (manufacturing techniques) - 54 forging
(manufacturing techniques) - 55 boiler metal work (manufacturing
techniques) - 56 sheet metal work (manufacturing techniques) - 57
general machining work - 58 precision machining work - 59 specialised
techniques - 61 vacuum pumps - 62 refrigeration equipment - 63
gas-handling equipment - 64 storage and transport of cryogens - 65
measurement equipment (vacuum and low- temperature technology) - 66
low-temperature materials - 67 vacuum components \& chambers - 68
low-temperature components - 69 vacuum and low-temperature technology -
71 films and emulsions - 72 scintillation counter components - 73 wire
chamber elements - 74 special detector components - 75 calorimeter
elements 8A radiation protection - n.a. not available). Center: CERN
external procurement - commitment for total and high-tech orders (pCp:
Past CERN procurement - commitment (k\euro\ 2013) tHp1: Total high-tech
procurement - commitment (k\euro\ 2013) tHp2: Total high-tech procurement
- commitment - only orders {\textgreater} 50 kCHF (ke2013)) (right);
distribution of EBITDA 2013 from ORBIS in firms at NACE industry levels
matched with CERN codes (right). Bottom: ROOT download data (left);
ENPV Cumulative distribution function conditional to PDF of critical
variables (k\euro\ 2013) (right).} Source: Authors'
elaboration of CERN data.}
\label{fig:tech}
\end{figure}

We have estimated the incremental profits based on LHC-related
procurement orders (categorized according to activity and technological
intensity codes), which we forecast up to 2025, and then used these
values to determine incremental turnover for the suppliers through
estimates of economic utility/sales ratios from Bianchi-Streit et al.
1984 and Autio et al. 2003 (based on surveys to CERN suppliers) and
EBITDA margins data (a measure of gross profits/sales ratio) for
companies in related sectors, extracted from the ORBIS database (BVD)
of companies balance sheets\footnote{ \url{https://orbis.bvdinfo.com}}
(see Fig.~\ref{fig:tech}). 

We explain here the estimation process in some detail. The total value
of CERN procurement by year and by activity code has been recovered
from the CERN Procurement and Industrial Services Companies (personal
communication, October 2013). A sample of 300 orders exceeding 
$10,000$ CHF in nominal value has been extracted from a data set
provided to us by the aforementioned CERN office. Each sampled order
has been classified (with the help of expert CERN staff) according to a
five-point scale: 1) {\textquotedblleft}very likely to be off-the-shelf
products with low technological intensity{\textquotedblright}; 2)
{\textquotedblleft}off-the-shelf products with an average technological
intensity{\textquotedblright}; 3) {\textquotedblleft}mostly
off-the-shelf products, usually high-tech and requiring some careful
specifications{\textquotedblright}; 4) {\textquotedblleft}high-tech
products with a moderate to high specification activity intensity to
customize product for LHC{\textquotedblright}; 5)
{\textquotedblleft}products at the frontier of technology with an
intensive customization work and co-design involving CERN
staff.{\textquotedblright} An average technological intensity score has
been attributed to each CERN activity code; we have classified as
high-tech the codes with average technological intensity class equal or
greater than 3. This led to the identification of 23 high-tech activity
codes. 

Procurement value has then been computed only for orders related to
these codes, which turned out to be 35\% of the total of procurement
expenditures. This would be only 17\% if we exclude orders below 50,
$000$ CHF and 58\% if we include orders below this threshold and for
other activity codes. We took a triangular distribution with average
and mode model equal to 35\% and minimum and maximum as the above
range. A share of 84\% of yearly total expenditures of Collaborations
is attributed to external procurement, using the same share as CERN.
This share has been used also for the future forecasts of both CERN and
the Collaborations up to 2025, based on the previous forecast of cost
trends. 

For the Collaborations, which are known to include a significantly
higher share of high-tech orders, we assume a triangular distribution
of the share of high-tech procurement with average and mode equal to
58\% and with minimum set to 40\% and maximum to 75\%, based on expert
assessment. We have then identified 1,480 benchmark firms from the
ORBIS database in the year 2013 and in six countries (Italy, France,
Germany, Switzerland, UK, USA). These countries were selected because
they received 78\% of the total CERN procurement expenditure between
1995 and 2013\footnote{ Data on procurement commitment by country
provided by CERN staff, October 2013.} . In selecting this sample, we
have considered companies whose primary activity matches with the
corresponding CERN activity codes\footnote{ The following NACE sectoral
codes have been considered: manufacture of basic metals (24);
manufacturing of structural metal products (25.1); forging, pressing,
stamping, and roll-forming of metal (25.5); manufacturing of other
fabricated metal products (25.9); manufacturing of computer,
electronic, and optical products (26); manufacturing of electrical
equipment (27); manufacturing of machinery and equipment not classified
elsewhere (28); specialised construction activities (43);
telecommunications (61); computer programming, consultancy, and related
activities (62); information service activities (63).}. After having
observed the EBITDA margin sample distribution, we have computed an
average (13.1\%) and standard deviation EBITDA, weighted by country,
and used these parameters to define a normal distribution of the
EBITDA. We have then estimated the incremental turnover over 5 years by
the LEP average utility/sales ratio to be equal to 3, based on the
results of Bianchi-Streit et al. 1984 and Autio et al. 2003, which in
turn are within the range of other studies as reported in Table 2.
Based on these sources, we assumed a triangular distribution with mode
equal to the mean, minimum 1.4, maximum 4.2. This ratio has been
applied to the high-tech procurement of both CERN and Collaborations.
We have finally computed the additional sales times EBITDA margin, thus
estimating the incremental profits of firms in the LHC supply chain in
other markets. 

Further benefits to businesses or organizations providing LHC services
come from software developed for analyzing the LHC experimental data
and made available for free: ROOT (about 25,000 users in 2013 outside
physics, mostly in the finance sector) and GEANT4 (used, e.g., in
medicine for simulating radiation damage in DNA). The benefits of the
externality are estimated as the avoided cost for the purchase of an
equivalent commercial software application (ROOT) or the cost required
for development of an analogous tool (GEANT4). The details of our
approach are as follows.

The number of ROOT users outside the high-energy physics community were
estimated on the basis of yearly download statistics of the software
code\footnote{\textstylefootnotereference{
}\url{https://root.cern.ch/drupal/content/download-statistics} } as
well as interviews and personal communications with CERN Physics
Department staff. We then forecast future trends based on
extrapolations of calibrated estimates of CERN staff on the basis of
past yearly downloads. This leads to a baseline forecast of 55,000
outside users in 2025. This has been taken as a stochastic variable
with a triangular distribution and a range of {\textpm}20\% about equal
average and mode. The number of new users by year has been estimated
based on data provided by CERN staff. The market prices of several
comparable commercial software codes have then been analyzed. The range
of avoided costs, depending on computing needs, goes from zero (if the
R open-source statistical analysis code was used instead) to 17~k\euro\
per year for a one-year license\footnote{\textstylefootnotereference{
}If, e.g., Oracle Advance Analytics was used.}. We have assumed a
triangular yearly cost-saving PDF for each ROOT user, with average and
mode equal to 1.5k\euro, minimum set to 1~k\euro\ and maximum to 2~k\euro. Based on interviews with experts, we have assumed a trapezoidal
PDF for the number of usage years, with two modes equal respectively to
3 and 10; minimum 0; maximum 20, based on actual data inspection. The
number of users, times the avoided cost per year, is then discounted
and summed to compute the PV of the ROOT-related benefit. 

For GEANT4\footnote{ http://geant4.web.cern.ch/geant4/license/} we have
identified about fifty research centers, space agencies, and firms in
which it is routinely used (not including a substantial number of
hospitals that use GEANT4 for medical applications). Out of this list,
we have made a distinction between the 38 centers that contributed in
some form to the development of the code versus the remaining ones. The
avoided cost is based on the production cost of GEANT4 (around 35~M\euro\ up to 2013, provided by
CERN staff and generated using SLOCcount\footnote{
\href{http://www.dwheeler.com/sloccount}{www.dwheeler.com/sloccount}};
the total CERN contribution to this cost is estimated to be 50\%. The
avoided cost for the aforementioned 38 centers is reduced to the
contribution they actually provided (assumed to be the same for each
centre, thus 50\% of 35~M\euro\ divided by 38), while it is the
full GEANT4 cost for the remaining ones. A forecast to 2025 and a
yearly avoided cost has then been estimated. The total cumulated
avoided cost has been taken as a symmetric triangular PDF {\textpm}30\%
about a mode and mean both equal to 2.8~G\euro. 

To sum up: the total mean value of the technological benefits is 
$\langle TE\rangle$, of which around 62\% arises from open software
and the remainder from incremental profits for firms because of sales
to customers other than CERN. The technological benefits pay back 39\%
of the total cost.

\begin{table}{\scriptsize
\tablehead{}
\begin{supertabular}{m{0.8948598in}m{1.3163599in}m{1.4163599in}m{2.74976in}}
\hline
\centering \textbf{\textcolor{black}{Average values}} &
\centering \textbf{\textcolor{black}{Research organization}} &
\centering \textbf{\textcolor{black}{Method of estimation}} &
\centering\arraybslash \textbf{\textcolor{black}{Source}}\\\hline
\centering \textbf{\textcolor{black}{3}} &
\centering \textcolor{black}{CERN} &
\centering \textcolor{black}{Survey of firms} &
\centering\arraybslash \textcolor{black}{Schmied (1975); }\\
\centering \textbf{\textcolor{black}{1.2}} &
\centering \textcolor{black}{CERN} &
\centering \textcolor{black}{Survey} &
\centering\arraybslash \textcolor{black}{Schmied (1982);}\\
\centering \textbf{\textcolor{black}{3}} &
\centering \textcolor{black}{CERN} &
\centering \textcolor{black}{Survey} &
\centering\arraybslash \textcolor{black}{Bianchi-Streit et al.
}\textcolor{black}{(1984) }\\
\centering \textbf{\textcolor{black}{3}} &
\centering \textcolor{black}{ESA} &
\centering \textcolor{black}{Survey of firms} &
\centering\arraybslash \textcolor{black}{Brendle et al. (1980) and Bach
et al. (1988)}\\
\centering \textbf{\textcolor{black}{1.5-1.6}} &
\centering \textcolor{black}{ESA} &
\centering \textcolor{black}{Survey} &
\centering\arraybslash \textcolor{black}{Schmied (1982);}\\
\centering \textbf{\textcolor{black}{4.5}} &
\centering \textcolor{black}{ESA} &
\centering \textcolor{black}{Survey} &
\centering\arraybslash \textcolor{black}{Danish Agency for Science
(2008)}\\
\centering \textbf{\textcolor{black}{2.1}} &
\centering \textcolor{black}{NASA (Space Programmes)} &
\centering \textcolor{black}{Input-Output model} &
\centering\arraybslash \textcolor{black}{Bezdek and Wendling (1992)}\\
\centering \textbf{\textcolor{black}{2-2.7}} &
\centering \textcolor{black}{INFN} &
\centering \textcolor{black}{Input-Output model} &
\centering\arraybslash \textcolor{black}{Salina (2006)}\\
\centering \textbf{\textcolor{black}{3.03}} &
\centering \textcolor{black}{John Innes Centre} &
\centering \textcolor{black}{Input-Output model} &
\centering\arraybslash \textcolor{black}{DTZ (2009)}\\\hline
\end{supertabular}
\caption{ Economic utility's ratios in the literature.
Source: authors based on cited sources}
}\end{table}

\section{Benefits to the general public: visits to LHC and other
direct cultural effects }

There are direct cultural benefits of the LHC to the general public
visiting CERN and taking advantage of its exhibitions, websites, and
outreach activities, including their impact on the media. The general
valuation criteria for these benefits has been the revealed preference
of the WTP, estimated in different ways. The details of our estimation
are as follows. 

The key social groups that we have considered are: (a) onsite CERN
visitors; (b) visitors to CERN travelling exhibitions; (c) people
reached by media reporting LHC-related news; (d) visitors to CERN and
Collaborations websites; (e) users of LHC-related social media
(YouTube; Twitter; Facebook; Google+); (f) participants in two
volunteer computing programs. 

\begin{figure}
\includegraphics[width=6.6929in,height=4.0646in]{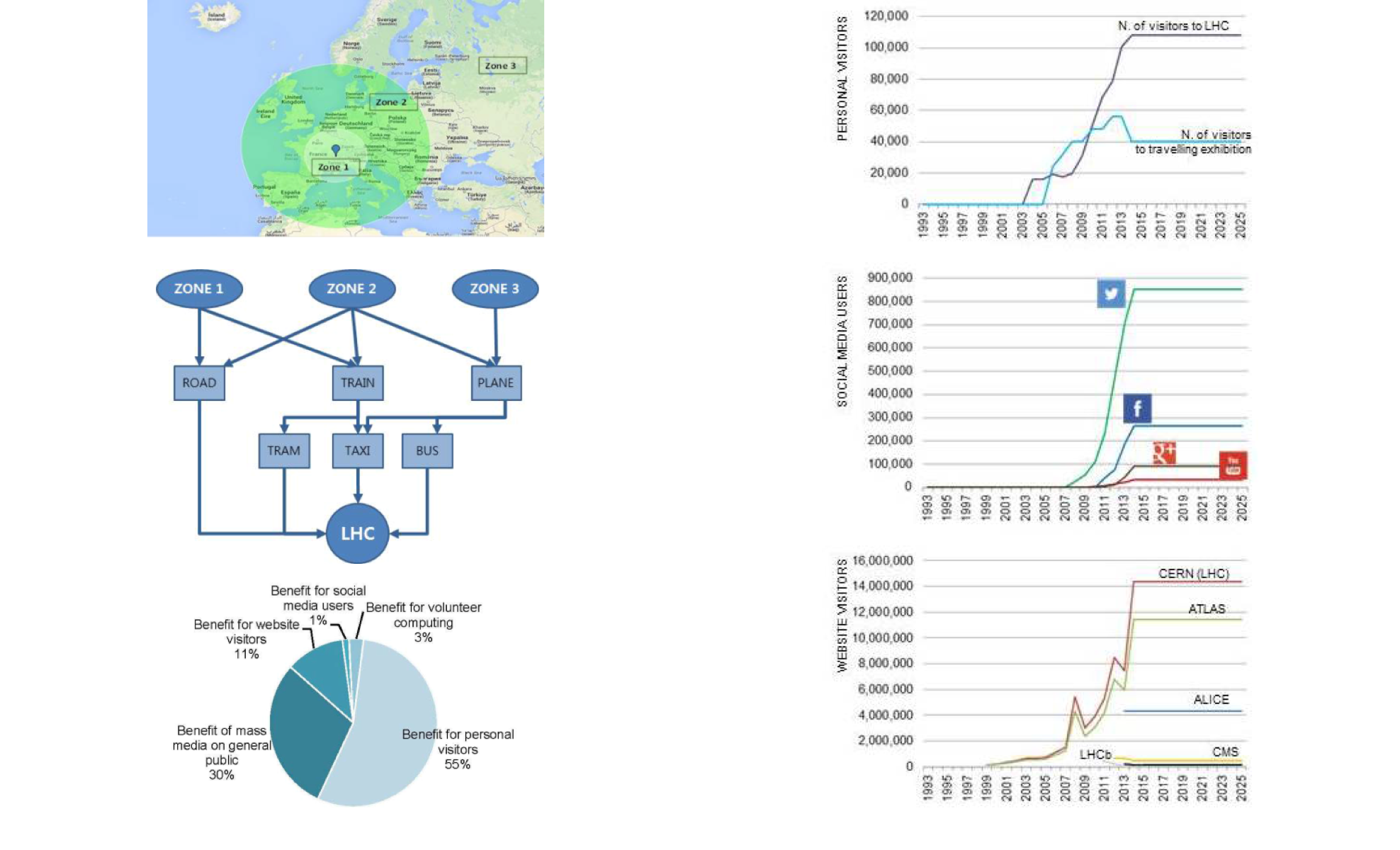}
\caption{ Left: (from top to bottom) Travel zones for CERN for
visitors; CERN visitors by mode of transport; share of benefits by type
of outreach activity (Cumulated impact to 2025). Right: benefits to
personal, visitors, social media users and website visitors.
The valuation of the benefit is based on the segmentation of visitors in
three areas of origin with increasing distance from CERN (see Fig.~\ref{fig:tour}),
and by average travel costs for each zone, based on seven origin cities
taken as cost benchmarks. For each zone, a transport mode combination
and length of stay have been assumed (see Fig.~\ref{fig:tour}). The three zones and
the share of visitors for each zone are based on data provided by the
CERN Communication Group (personal communication October 2013);
additional costs have been estimated including for accommodation and
meals (data extracted from the CERN website).}
\label{fig:tour}
\end{figure}

(a) Benefits for on-site visitors are determined using the revealed
preference method (Clawson and Knetsch 1966), with the MSV of the time
spent in travelling obtained from HEATCO\footnote{
\url{http://heatco.ier.uni-stuttgart.de/}} data (see Fig.~\ref{fig:tour}). Data for
onsite visitors since 2004 to 2013 have been provided to us by the
Communication Groups of CERN and by each Collaboration. The forecast to
2025 has been extrapolated by a constant yearly value, based on the
trend observed in the previous years. We have estimated an 80\% overlap
between visitors to LHC experiment facilities and the permanent CERN
Exhibitions (Microcosm and Universe of Particles in the Globe of
Science and Innovation); moreover, only 80\% of visitors to CERN have
been attributed to the LHC. 

 The value of travelers{\textquoteright} time is based on HEATCO for
each member state and for some non-members. Based on the distribution
of visitors by country and mode of transportation, we have estimated an
overall distribution of visitors based on the following PDF: trapezoid
distribution for air travelers (minimum equal to 5; maximum equal to
45, first mode equal to 22 and second mode equal to 27, all in
\euro/hour: there are two modes because of the difference between the two
main origin groups and this suggests using a trapezoid PDF); triangular
distribution for travel by car and train (mean and mode equal to 18;
minimum 6 and maximum 30). 

(b) For the CERN travelling exhibitions, we have used the number of past
visitors as provided by CERN (between 30,000 and 70,000 for the period
2006-2013). We have assumed a constant number of 40,000 visitors per
year during from 2014 to 2035. The WTP is prudentially assumed to be
just 1~\euro\ per visitor (assuming local transport). 

(c) For the benefit of LHC coverage in the media, we have conservatively
considered only the news spikes on September 10 2008 (first run of LHC)
and July 4 2012 (announcement of the discovery of the Higgs
boson)\footnote{ Sources for these point estimates are: New Scientist
(2008) and \url{http://cds.cern.ch/journal/CERNBulletin/2012/30/News
Articles/1462248}}. We have estimated, based on some interviews, that
the average time devoted to each LHC news per head is 2 minutes. We
have treated the audience number as a stochastic variable, assuming a
triangular distribution (minimum zero, maximum one billion, average and
mode equal to 0.5 billion). The value of time of the target audience
has been estimated based on current GDP per capita in the average CERN
Member States and the USA (for 2013, using IMF data), and the number of
working days per year (8 hours times 225 working days). This is treated
as a stochastic, triangular distribution, with minimum equal to 3~\euro;
maximum 42~\euro, and mode and mean equal to 17~\euro. 

(d) We have estimated the number of website visitors on the basis of
historical data on hits until 2013-2104 (source CERN and Communication
Groups in the main Collaborations). Our forecast is conservatively
based in assuming that the value at the last available observation
remains constant. The benefit comes from the number of minutes per hit
from users of the websites, estimated to be a triangular distribution
with average and mode equal to 2 minutes, and ranging from 0 to 4
minutes. 

(e) Further benefits come from LHC-related social media and website
visits, with the MSV of time of the general public proxied by the
hourly value of per capita GDP (see Fig.~\ref{fig:tour}). For social media usage, we
recovered data provided by CERN and Collaborations, attributing to the
LHC 80\% of the hits to CERN-related social media and 100\% of those
related to the Collaborations. We used historical data until 2014 and
for the subsequent years we have taken the last year{\textquoteright}s
data as constant. The average stay time is assumed for all social media
to be distributed according to a triangular distribution with average
and mode equal to 0.5 minutes per capita, ranging from zero to one
minute. Time is then valued as above. 

(f) Finally, some CERN projects exploit computing time donated from
volunteers to run simulation of particle collisions, with WTP revealed
by time spent. Two such LHC-related programs are SIXTRACK and
TEST4THEORY, where outsiders donate to CERN the machine time and
capacity of their own computers and are then able to access some data
and to join a social network. The stock number of volunteers in 2013
has been provided by the CERN PH Department (personal communication);
based on this information, we have assumed a rate of increase from the
program start years (respectively 2007 and 2001). A forecast of the
future volunteer stock has been given to us to 2025 by the same source;
again, we have assumed a yearly rate of change over the years
2014-2025. The opportunity cost is the time to download, install, and
configure the programs (15 minutes per capita \textit{una tantum}) and
the time spent in forum discussions (15 minutes per month per capita).
Again, time is valued as above.

The total mean value of the above mentioned cultural effects is 
$\left\langle CU\right\rangle = 2.1$~G\euro.
This value contributes around 16\% against the total cost.

\section{Non-use benefits: scientific knowledge as a public good}

As mentioned in section 2, beyond the direct benefits accruing to
certain social groups, there is a non-rival and non-excludable benefit,
i.e., a public good arising from the LHC{\textquoteright}s discoveries.
This is not connected to any specific use of such discoveries, but only
to the social preference for knowing that such new knowledge will be
available; this is a non-use value:

{\textquotedblleft}A resource or a service might be valued even if it is
not consumed. Such values are referred to as \textit{non-use values},
bur sometimes they are labeled \textit{passive-use} or
\textit{intrinsic values. }{\dots}\textit{ }If the project being
evaluated affects non-use values this should be reflected in the
cost-benefit analysis~{\dots}~among these are \textit{existence
values}{\textquotedblright} (Johansson and Kristr\"om, 2015, pp 24-25).

The empirical estimation of non-use value in environmental and cultural
economics is generally based on contingent valuation approaches and
their variants. The issue is discussed in some detail in Florio and
Sirtori (2015). The benchmark methodology in the literature is the NOAA
1993 panel (Arrow et al. 1993), but there have been several advances
since then (see Carson 2012 for a review and Johansson and Kristr\"om,
Ch. 9). We wish to determine social preferences for the non-use value
of LHC discoveries, a public good with yet unknown practical use. 

We have thus designed a contingent valuation study tailored to our
problem. Ideally, a random sample of taxpayers in the CERN Member
States and in other countries (e.g., notably the USA) supporting the
LHC in different forms would be needed. An in-depth survey, as
recommended by the NOAA panel, needs personal interviews of a
representative sample of the population, but in our case spreading a
manageable sample across many countries and types of individuals would
be too costly and not necessarily more reliable than performing a more
focused survey. Hence, we have targeted university students for
in-depth personal interviews in four CERN Member States as
representative of future taxpayers with tertiary education. Referring
to students in experimental economics and political science is common
practice (see, e.g., Druckman and Kam 2011). In fact, to be
conservative, we have assumed that all taxpayers with less than
tertiary education would be willing to contribute nothing to scientific
discovery by the LHC as a pure public good. Surely, in this way, we
grossly understate the social preferences, as at least some people with
less than tertiary education may have a positive WTP. The results were
used to guess the WTP of taxpayers with tertiary education in CERN
Member States and from non-Member States.

\begin{figure}
\includegraphics[width=6.6929in,height=1.9437in]{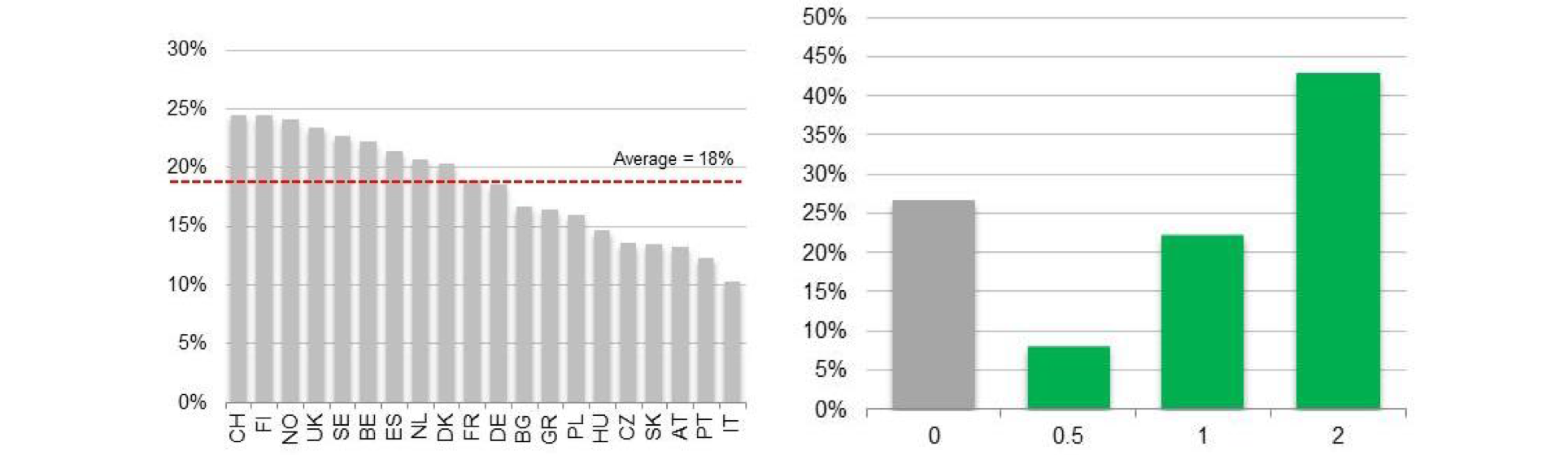}
\caption{Share of adult population (18-74 years old) with at
least tertiary education (left); average annual WPT of the respondents
to the survey (right).}
\label{fig:coval}
\end{figure}
This survey on WTP for the LHC-related public good was undertaken in
Milan in October-November 2014 and in Exeter (UK), Paris (France), and
A Coru\~na (Spain) in February-March 2015: 1027 questionnaires were
collected. The average time spent answering the questionnaire (28
questions) was about 25 minutes\footnote{ For the questionnaire and
other details, see Catalano et al. 2016. }. The respondent was first
given a one-page summary of the LHC Wikipedia page as an information
set. The geographical distribution of respondents was 40\% from Italy
and 20\% each from Spain, France, and the UK. Out of the total number
of respondents, 85\% were 19-25 years old, while the remainder were
more than 26 years old. Out of the respondents 57\% were females. A
share of 64\% were in the humanities and social sciences, with the
remainder in science-related curricula. Questions included: household
composition, family income, personal income, high-school background,
previous knowledge of research infrastructure, source of information,
if any, on the LHC and the Higgs boson discovery, whether the
respondent has ever visited CERN, interest in science, willingness to
pay for LHC research activities a fixed lump-sum or a yearly economic
contribution over 30 years, in pre-set discrete amounts (zero, 0.5, 1,
2~\euro)\footnote{ As we did not consider a higher range of values, this
truncates the right tail of the distribution in such a way that in fact
we are underestimating the WTP. To double-check the preferences, the
questionnaire included also a question on the WTP a lump sum
contribution of 30 euros in a
{\textquotedblleft}referendum-like{\textquotedblright} format, as
recommended by the NOAA panel and in Catalano et al. (2016); we use
this alternative {\textquotedblleft}referendum{\textquotedblright}
question format to double-check the results reported here. }.  Only
answers to the last question (yearly contribution) are used here, while
all the other variables have been used for a detailed statistical
analysis by Catalano et al (2016). 

We have then taken the sample average yearly WTP, weighted by the number
of respondents by country, for only those respondents who declared a
positive annual WTP, these comprising 73\% of the total. This has given
us a sample distribution with three discrete values (0.5, 1, and 2~\euro), and mode and maximum equal to 2. Each annual WTP has then been
multiplied (undiscounted, as this is an instant variable) by 30 years.
This per capita WTP has been applied to 73\% of respondents between
18-74 years of age with at least tertiary education coming from CERN
Member States, determined by data from Eurostat 2013 (see Fig.~\ref{fig:coval}). We have then
added to the previous target population an additional 21\% from CERN
non-member states, reflecting the share of onsite visitors to CERN from
non-member states (visitor statistics provided by CERN staff as a
personal communication). We have treated the per capita WTP as a
stochastic variable, assuming a truncated triangular probability
distribution with maximum and mode equal to 2~\euro\ and minimum equal to
0.1~\euro, reflecting the sample distribution for non-zero values. 

The undiscounted mean non-use value is found to be  $\left\langle
EXV_{0}\right\rangle = 3.2$~G\euro, paying
back around 24\% of total costs.

\section{Summary of results and concluding remarks}

Based on the forecasts of social costs and benefits in the previous
sections, we have determined the probability distribution of the net
present value of the LHC as for Eq.~(\ref{eq:npvres}) by running a Monte Carlo
simulation (10,000 draws conditional to the PDF of the nineteen
stochastic variables mentioned
above\footnote{\textstylefootnotereference{ }The full list and details
of the simulations are available upon request.}). Each draw generates
an $NPV$ estimate in a state of the world supported by a random set of
the possible values taken by the model stochastic variables. The number
of variables we have considered for the Monte Carlo simulation and the
number of draws are largely in excess of what is usually done in the
evaluation of large-scale investment projects by international and
national bodies (Florio 2014, OECD 2015), e.g., for high-speed rail
infrastructure that faces considerable uncertainty and optimism bias
(Flyvbjerg et al. 2003). While we have been prudent, and even
pessimistic, in our assumptions, caution is necessary in the
interpretation of the final results, which we will briefly summarize
and discuss here. As with any forecast covering the long run, there is
obviously some residual uncertainty, but we are confident that residual
estimation errors are mostly in the direction of underestimating the
net social benefit of the LHC. This was deliberate, as we have
preferred to be conservative.

The total present value to 2025 of operating and capital expenditure of
the LHC is estimated at 13.5~G\euro\ (net of the cost of scientific
personnel). In terms of contributions to the sum of the social benefits
(16.4~G\euro), the present value of the human capital effects and
of technological spillovers are the most important ones, and of similar
size, each contributing around one third of the benefits. Adding the
tiny secondary effect of the publications (net of the direct value of
LHC research outputs), around 68\% of the socio-economic benefits is
related to professional activities (within firms, academia, and other
organizations), while the remaining benefits spill over to the general
public, either as a direct cultural effect (a private good) or as a
pure public good (a non-use benefit). Any other (if any) unpredictable
social benefits of future applications of scientific discoveries at the
LHC are excluded from our analysis; they will remain as an extra bonus
for future generations, donated to them by current taxpayers.

\begin{figure}
\includegraphics[width=6.6929in,height=1.8744in]{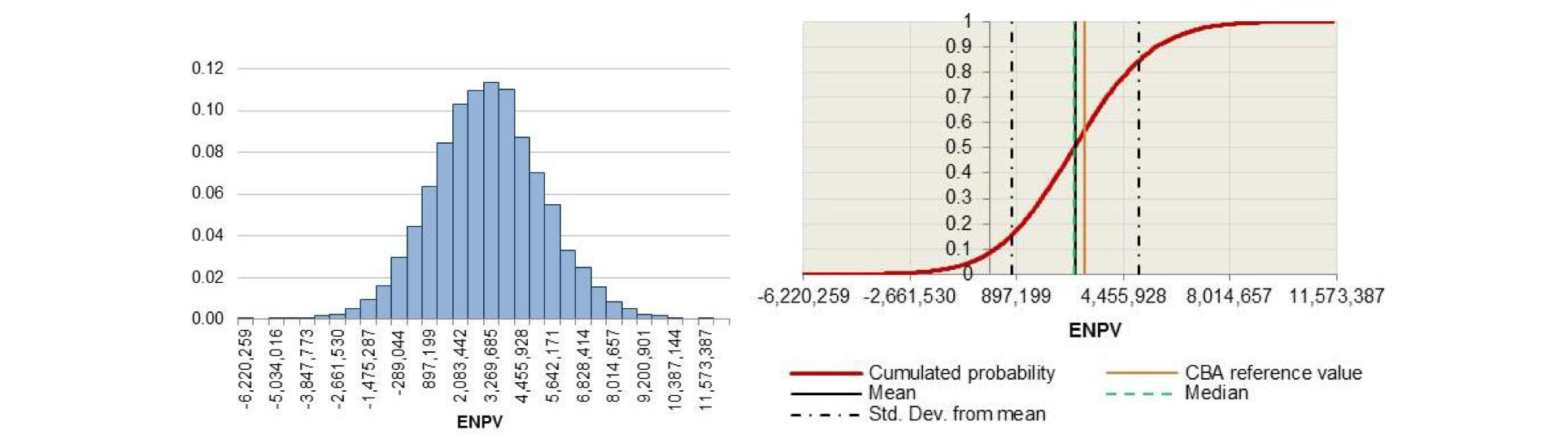}
\caption{ Net present value PDF (left) and cumulative
distribution (right).}
\label{fig:npv}
\end{figure}

The final PDF and cumulative probability distribution for the $NPV$ are
shown in Fig.~\ref{fig:npv}. We find that the expected $NPV$ of the LHC is around 2.9~G\euro, with a conditional probability of a negative $NPV$ smaller
than 9\% with a $3\sigma $ Monte Carlo error below 2\%. The expected
benefit/cost ratio is around 1.2 and the expected internal rate of
return is 4.7\%~\footnote{ The $NPV$ would be lower if an opportunity
cost of public funds is considered because of distortionary taxation,
but it would still be positive for the typical current range in
developed countries. In fact, European Commission (2014) does not
recommend introducing a correction for the opportunity cost of public
funds for projects funded by grants supported by international
transfers, because it would not be clear which is the relevant source
of funding. For example, if it is sovereign debt in Europe, for most of
the core Member States of the CERN the real interest rate on such debt
for 30 years bonds would be largely below the social discount rate that
we use and below the long-term rate of growth of GDP. A sensitivity
analysis can ascertain the relative impact of lowering the social
discount rate and of increasing the total cost by a correction factor
related to the opportunity cost of public funds, but we leave this and
other sensitivity analysis issues for further research. }.

We have thus shown how a social CBA probabilistic model can be applied
to evaluate a large-scale research infrastructure project, based on
empirical methods. The main novelty of this contribution is that we
show the feasibility, following the Florio and Sirtori (2015) approach,
of a quantitative valuation of the socio-economic impact of such
infrastructure in a way consistent with first principles of applied
welfare economics. The way we respectively define and apply the
distinction between the use and non-use benefits of research
infrastructures, and of the measurable and non-measurable impacts, are
also novel relative to the previous literature, as discussed in Florio
and Sirtori (2015). Moreover, our treatment of risk, while based on
standard Montecarlo methods, shows a way to forecast some stochastic
variables typical of a CBA model of large scale RI.

Clearly the LHC is a special, albeit important case of an RI, because of
the long time of construction and operation, the high number of
scientists, students and post-docs involved, the large number of firms
in the supply chain, the externalities from the open access to
software, the wide coverage in the media and the attraction of onsite
visitors, and the nature of a frontier basic research facility.
However, we believe that the role of a case study in social science is
to suggest new avenues of inquiry. As stated by Flyvbjerg (2006 p.219):

{\textquotedblleft}A scientific discipline without a large number of
thoroughly executed case studies is a discipline without systematic
production of exemplars, and that a discipline without exemplars is an
ineffective one.{\textquotedblright} 

 It would hence be necessary to expand further the evaluation of the
socio-economic impact of RIs to other large-scale facilities, including
those in applied science. An example of the latter is a recent study of
the CNAO particle accelerator for hadron therapy (Pancotti et al.
2015), which uses, in the context of medical research, the same
methodology we apply here. The proportion and scale of the costs and
benefits may be different elsewhere, but we believe that the main
ingredients of a CBA of research infrastructure are well represented in
the LHC case; hence replication can be attempted if data are available.
Further studies on a range of different facilities, in different
science and technology domains, and in different countries, are needed
to confirm our intuition. 

\textbf{Acknowledgements:} We are grateful to two anonymous reviewers
for helpful criticism and advice, and for comments on earlier versions
of the manuscript to Antonella Calvia Gotz (EIB), Albert De Roeck
(CERN), Andr\'es Fai\~na (University A Coru\~na), Anna Giunta
(University of Rome III), Ugo Finzi (CSIL), Johannes Gutleber (CERN),
Diana Hicks (Georgia Institute of Technology), Per-Olov Johansson
(Stockholm School of Economics), Mark Mawhinney (EIB/JASPERS), Giorgio
Rossi (ESFRI), Lucio Rossi (CERN), Herwig Franz Schopper (CERN Director
General Emeritus), Florian Sonneman (CERN), Alessandro Sterlacchini
(Universit\`a Politecnica delle Marche), Anders Unnervik (CERN), Witold
Willak (European Commission) and many others, including participants in
the XIII Milan European Economy Workshop (12-13 June 2014), the Annual
Conference of the Benefit-Cost Society (Washington DC 12-20 March
2015), the CERN Colloquium (Geneva 11 June 2015), the workshop of the
OECD Global Science Forum on socio-economic impact of research
infrastructures (Paris November 3, 2015), the final UNIMI-EIBURS
workshop (Brussels November 13, 2015), the European Space Agency
meeting of the Space Economy Steering Committee (Paris November, 25
2015). S.F. thanks T.~Camporesi, S.~Carrazza and G.~Giudice for
illuminating discussions.

This paper has been produced in the frame of the project Cost-Benefit
Analysis in the Research, Development and Innovation Sector sponsored
by the EIB University research programme (EIBURS). The findings,
interpretations and conclusions presented in the paper should not be
attributed to the EIB or any other institutions. The work of S.F. is
partly supported by an Italian PRIN2010 grant and by the Executive
Research Agency (REA) of the European 
Commission under the Grant Agreement PITN-GA-2012-316704 (HiggsTools).

\textbf{References}

Abt, H.A. and E. Garfield. 2002. Is the Relationship between Numbers of
References and Paper Lenghts the Same for All Sciences? Journal of the
American Society for Information Science and Technology, 53\textbf{,}
1106-1112. 

Arrow, K. J., and A.C. Fisher. 1974. Environmental Preservation,
Uncertainty, and Irreversibility, Quarterly Journal of
Economics,\textbf{ }88, 312-19.

Arrow, K., R. Solow, P.R. Portney, E.E. Leamer, R. Radner and Schuman,
H. 1993. Report on the NOAA panel on contingent valuation. NOAA.

Atkinson, G. and S. Mourato. 2008. Environmental Cost-Benefit Analysis.
Annual Review of Environment and Resources. 33, 317-344.

Autio, E., M. Bianchi-Streit and A.P Hameri A.P. 2003. Technology
transfer and technological learn- ing through CERNs procurement
activity. CERN-2003-005.

Bacchiocchi, E. and F. Montobbio, F. 2009. Knowledge diffusion from
university and public re- search. A comparison between US, Japan and
Europe using patent citations. Journal of Technology Transfer, 34,
169-18.

Bach L., G. Lambert, S. Ret, J. Shachar, with the collaboration of R.
Risser, E. Zuscovitch, under the direction of P. Cohendet, M-J. Ledoux.
1988. Study of the economic effects of European space expenditure, ESA
Contract No. 7062/87/F/RD/(SC).

Baum, W.C. and S.M. Tolbert. 1985. Investing in Development: Lessons of
World Bank Experience Oxford: Oxford University Press.

Bezdek, R.H. and R.M. Wendling. 1992. Sharing out NASA{\textquoteright}s
spoils. Nature, 355(6356), 105-106.

Bianchi-Streit, M., N. Blackburne, R. Budde, H. Reitz, B. Sagnell, H.
Schmied and B. Schorr. 1984. Economic utility resulting from CERN
contracts (second study), CERN, 84-14.

Boardman, A.E., D.H. Greenberg, A.R. Vining, and D.L. Weimer. 2006.
Cost-Benefit Analysis: Concepts and Practice. Upper Saddle River NJ:
Pearson Prentice Hall.

Boisot, M., M. Nordberg, S. Yami and B. Nicquevert (eds). 2011.
Collisions and Collaboration. The Organization of Learning in the ATLAS
experiment at the LHC. Oxford: Oxford University Press.

Brendle P., P. Cohendet, J.A. Heraud, R. Larue de Tournemine, H.
Schmied. 1980. Les effets \'economiques induits de l{\textquoteright}
ESA. ESA Contracts Report Vol. 3.

Camporesi, T. 2001. High-energy physics as a career springboard,
European Journal of Physics, 22, 2, 159-148.

Carrazza, S., A. Ferrara and S. Salini. 2014. Research infrastructures
in the LHC era: a scientometric approach. Working Paper DEMM 2014-12,
University of Milan.

Carson, R.T. 2012. {\textquoteleft}Contingent Valuation: A Practical
Alternative when Prices Aren{\textquoteright}t
Available{\textquoteright}, Journal of Economic Perspectives, 26(4),
27-42.

Catalano G, C. Del Bo, M. Florio. 2015a. Human Capital Formation at LHC:
Survey Results.
\url{http://wp.demm.unimi.it/tl_files/wp/2015/DEMM-2015_10wp}. 

Catalano G, Florio M., Giffoni F,. Forthcoming. Contingent valuation of
social preference for science as a pure public good: the LHC case. DEMM
Working paper, University of Milan. 

CERN Communication Group. 2009. LHC the Guide FAQ.
\url{http://cds.cern.ch/record/1165534/files/CERN-Brochure-2009-003-Eng.pdf}

Clawson, M. and J.L. Knetsch, J.L. 1966. Economics of outdoor
recreation. Baltimore: Johns Hopkins Press.

Danish Agency for Science. 2008. Evaluation of Danish industrial
activities in the European space agency (ESA). Assessment of the
economic impacts of the Danish ESA membership. Technical report, Danish
Agency for Science.

Dr\`eze, J. and N. Stern. 1990. Policy reform, shadow prices and market
prices. Journal of Public Economics 42, 1-45. 

Dr\`eze, J. and N. Stern. 1987. The theory of cost-benefit analysis. In
A.J. Auerbach and M. Feldstein (eds), \textit{Handbook of public
economics}. Vol. 2, pp. 909-989. Amsterdam: North Holland.  

Druckman, J N and C.D Kam, 2011 Students as experimental participants. A
defense of the narrow data base, in James N. Druckman, Donald P. Green,
James H. Kuklinsi, and Arthur Lupia eds, Cambridge Handbookof
Experimental Political Science, New York: Cambridge University Press

DTZ. 2009. {\textquoteleft}Economic impact of the John Innes
Centre{\textquoteright}. Technical report, DTZ, Edinburgh.

Eckhardt, R. 1987. Stan Ulam, John von Neumann, and the Monte Carlo
method, Los Alamos Science 15, 131-137.

European Commission. 2014. Guide to Cost-Benefit Analysis of Investment
projects. DG Regional and Urban Policy.

European Investment Bank. 2013. The Economic Appraisal of Investment
projects at the EIB. 

Feller, I., C.P. Ailes and J.D. Roessner. 2002. Impacts of research
universities on technological innovation in industry: evidence from
Engineering Research Centers. Research Policy, 31, pp.457- 474. 

Florio, M. 2014. Applied Welfare Economics: Cost-Benefit Analysis of
Projects and Policies. New York: Routledge.

Florio, M. and E. Sirtori. 2015. Social Benefits and Costs of Large
Scale Research Infrastructures, Technology Forecasting and Social
Change, \url{http://.doi/10.1016/j.techfore.2015.11.024}

Florio, M., Forte, S., Pancotti, C., Sirtori, E., Vignetti, S. 2016.
Exploring Cost-Benefit Analysis of Research, Development and Innovation
Infrastructures: An Evaluation Framework. Working Papers, Centre for
Industrial Studies (CSIL).

Flyvbjerg, B. Buzelius, N. Rothengatter, W. 2003. Megaprojects and Risk:
An Anatomy of Ambition. Cambridge:
\href{https://en.wikipedia.org/wiki/Cambridge_University_Press}{Cambridge
University Press}. 

Flyvberg, B. 2006. Five Misunderstandigs about Case-Study Research,
Qualitative Inquiry, vol 12., n. 2, April pp 219-245. 

Harmon, C. 2011. Economic returns to education:what we know, what we
don{\textquoteright}t know, and where we are going. Some brief
pointers, IZA Policy Paper n.29, August.

Myles, G.D. 1995. Public Economics. Cambridge, MA: Cambridge University
Press

Hutton G. and G. Rehfuess. 2006. Guidelines for conducting cost-benefit
analysis of household energy and health interventions. WHO Publishing.

Johansson, P.O. 1991. An Introduction to Modern Welfare Economics.
Cambridge: Cambridge University Press. 

Johansson, P.O. 1995. Cost-Benefit Analysis of Environmental Change.
Cambridge: Cambridge University Press.

Johansson, P.O. and Kristrom B, 2015 Cost-Benefit Analysis for Project
Appraisal, Cambridge: Cambridge University Press.

Knight, F. 1921. Risk, Uncertainty and Profit. 2\textsuperscript{nd}
edition 1964. Tallahassee, FL: Sentry Press. 

Martin, B. R., and P. Tang. 2007. The benefits from Publicly Funded
Research, SPRU Electronic Working Paper Series, Paper no. 161.
University of Sussex, Brighton, UK: SPRU.

Montenegros C.E., Patrinos H.A. 2014, Comparable estimates of returns to
schooling around the world, Policy Research Working paper 7020,
September, Education Global Practice Group, World Bank Group

OECD. 2014. Report on the Impacts of Large Research Infrastructures on
Economic Innovation and on Society. Case studies at CERN. Global
Science Forum report. OECD Publishing, Paris. 

OECD. 2015. Government at a glance. Paris: OECD.

Pancotti, C., G. Battistoni, M. Genco, M.A. Livraga, P. Mella, S. Rossi
and S. Vignetti. 2015. The socio-economic impact of the National Hadron
therapy Centre for Cancer Treatment (CNAO): applying a cost-benefit
analysis analytical framework. DEMM Working Paper n. 2015-05.

Pearce, D.W., G. Atkinson, and S. Mourato. 2006. Cost-Benefit Analysis
and the Environment: Recent Developments. Paris: OECD.

Pouliquen, L.Y. 1970. Risk Analysis in Project Appraisal, World Bank
Staff Occasional Papers, 11. Baltimore: Johns Hopkins University Press.

Salina, G. 2006. Dalla ricerca di base al trasferimento tecnologico:
impatto dell{\textquoteright}attivit\`a scientifica
dell{\textquoteright}istituto nazionale di fisica nucleare
sull{\textquoteright}industria italiana, Rivista di cultura e politica
scientifica, No. 2/2006.

Salling, K. B. and S. Leleur S. 2011. Transport Appraisal and Monte
Carlo simulation by use of the CBA-DK model, Transport Policy,\textbf{
}18, 236-245.

Schmied, H. 1975. A study of economic utility resulting from CERN
contracts. Technical report, European Organization for Nuclear
Research, Geneva (Switzerland).

Schmied, H. 1982. {\textquoteleft}Results of attempts to quantify the
secondary economic effects generated by big research
centers{\textquoteright}. Engineering Management, IEEE Transactions on,
EM-29(4):154-165.

Schopper, H. 2009. The Lord of the Collider Rings at CERN, 1980-2000.
Heidelberg: Springer.

World Bank 2010, Cost-Benefit Analysis in World Bank projects, Fast
Track Briefs, June 4 2010, World Bank, Washington DC

\end{document}